\documentclass[submission,Phys]{SciPost}
\usepackage{color,amsmath,amssymb,multirow,hyperref,booktabs,graphicx,mathtools,xspace,hhline} 

\def\ptt{p_{T,t}^\text{truth}}
\def\mse{\sqrt{\text{MSE}}}

\newcommand{\kl}{\text{KL}}

\newcommand\one{\leavevmode\hbox{\small1\normalsize\kern-.33em1}}

\newcommand{\gev}{\text{GeV}}

\def\slashchar#1{\setbox0=\hbox{$#1$}           
   \dimen0=\wd0                                 
   \setbox1=\hbox{/} \dimen1=\wd1               
   \ifdim\dimen0>\dimen1                        
      \rlap{\hbox to \dimen0{\hfil/\hfil}}      
      #1                                        
   \else                                        
      \rlap{\hbox to \dimen1{\hfil$#1$\hfil}}   
      /                                         
   \fi}

\newcommand{\ie}{\textsl{i.e.}\;}

\newcommand{\be}{\begin{eqnarray*}}
\newcommand{\ee}{\end{eqnarray*}}

\newcommand{\bee}{\begin{eqnarray}}
\newcommand{\eee}{\end{eqnarray}}
\newcommand{\beeq}{\begin{equation}}
\newcommand{\eeeq}{\end{equation}}

\begin{document}

\begin{center}{\Large \textbf{
Per-Object Systematics using Deep-Learned Calibration
}}\end{center}

\begin{center}
Gregor Kasieczka\textsuperscript{1},
Michel Luchmann\textsuperscript{2}, 
Florian Otterpohl\textsuperscript{1}, and
Tilman Plehn\textsuperscript{2}
\end{center}

\begin{center}
{\bf 1} Institut f\"ur Experimentalphysik, Universit\"at Hamburg, Germany \\
{\bf 2} Institut f\"ur Theoretische Physik, Universit\"at Heidelberg, Germany \\
luchmann@stud.uni-heidelberg.de
\end{center}



\section*{Abstract}
{\bf We show how to treat systematic uncertainties using Bayesian deep
  networks for regression. First, we analyze how these networks
  separately trace statistical and systematic uncertainties on the
  momenta of boosted top quarks forming fat jets. Next, we propose a
  novel calibration procedure by training on labels and their error bars.
  Again, the network cleanly separates the different uncertainties. As
  a technical side effect, we show how Bayesian networks can be extended to
  describe non-Gaussian features.}

\vspace{10pt}
\noindent\rule{\textwidth}{1pt}
\tableofcontents\thispagestyle{fancy}
\noindent\rule{\textwidth}{1pt}
\vspace{10pt}

\newpage
\section{Introduction}
\label{sec:intro}

Modern methods of machine learning are becoming a crucial tool in
experimental and theoretical particle physics.  An especially active
field in this direction is subjet physics and jet
tagging~\cite{early_stuff}, where multi-variate analyses of high-level
observables are being replaced with deep neural networks working on
low-level inputs.  Early applications of deep learning techniques in
LHC physics rely on image recognition of jet
images~\cite{jet_images,jet_images2}. Their main challenge is to
combine calorimeter and tracking information, motivating graph
convolutional networks and point
clouds~\cite{particlenet}. Established benchmarks processes for these
methods include quark-gluon
discrimination~\cite{jets_qg,jets_qg2,jets_qg3,jets_qg4,jets_qg5,lola_qg},
flavor tagging~\cite{DeepJet},
$W$-tagging~\cite{jets_w,jets_w2,CMSTop,ATLASTop},
Higgs-tagging~\cite{jets_h,jets_h2}, or
top-tagging~\cite{deep_top1,deep_top2,lola,jets_top,jets_top2,jets_top3,jets_top4,jets_top5,CMSTop,ATLASTop}.
By now we can consider top jet classification at the level of tagging
performance as essentially solved~\cite{ml_review,jets_comparison}. This 
gives us room to consider question beyond the performance, for instance
what the networks are learning, how they can be visualized, how robust they
are, how we can control the uncertainties, and how machine learning methods
affect typical LHC analyses structurally.

One open question is driven by particle physics' obsession with error
bars: how do we quantify the different uncertainties in analyses using
neural networks~\cite{ours,ben_unc,bad,aussies}? This question is
related to visualization~\cite{capsules}, understanding the relevant
physics features~\cite{information,information2,information3,information4,information5}, and weakly supervised learning
approaches~\cite{weak,weak2,weak3,weak4,weak5,weak6,weak7,weak8} --- all combined under the general theme of
explainable AI. In LHC physics we have the advantage of excellent
Monte Carlo simulations and full control of the experimental
setup. This allows us to define and control different sources
uncertainties very precisely. If we accept that a neural network is
just a function relating training data to an output there exist (at
least) two main kinds of uncertainties:
\begin{enumerate} 
\item first, labelled training data comes with statistical and
  systematic uncertainties, where we define the former as
  uncertainties which vanish with more training data. The
  systematic uncertainties can be Gaussian or include shifts,
  depending on their sources. Unstable network training also belongs
  to this category of training-induced uncertainties~\cite{ours};
\item second, on the test data or analysis side we also encounter
  statistical and systematic uncertainties. When we include an
  inference or any kind of analysis we also encounter model or theory
  uncertainties~\cite{ben_unc}. For these uncertainties it is crucial
  that we ensure our analysis outcome is conservative.
\end{enumerate}
In a previous paper~\cite{ours} we have shown how Bayesian
classification networks can track uncertainties and provide jet-by-jet
error bars for the tagging output. Such a Bayesian network can
supplement a probabilistic classification output of `60\% signal' with
an error estimate of the kind `($60 \pm 10$)\% signal' for a given
jet. This kind of jet-by-jet information exceeds what is available
from standard LHC classification tools.  In principle, this approach
covers both, statistical errors from the size of the training sample
and systematic uncertainties for instance from the calibration of the
training sample. However, our quantitative analysis of Bayesian top
taggers encountered practical limitations, for instance that the jet
energy scale simultaneously affects the central value and the error
bar of the probabilistic output. A similar study of uncertainties just
appeared for a matrix element regression task~\cite{badger}.

In this follow-up study we look at this problem from a slightly
different angle, now defining the \textsl{regression task} of
extracting the energy of a tagged top quark inside a fat jet. Again,
we translate statistical and systematic uncertainties from the
training sample to the test output. The Bayesian network, introduces
in Sec.~\ref{sec:bayes}, allows us to construct a per-jet probability
distribution function over possible top momenta, or $p(p_t | \text{fat
  jet})$. The main advantage of using the regression task as example
is that it does not enforce a closed interval for the network output
and hence removes the correlation between central value and error
estimate in the network output. We use this advantage to cleanly
separate effects from the finite size of the training sample and from
the stochastic nature of the training sample in Sec.~\ref{sec:stat}.

In Sec.~\ref{sec:syst} the stochastic uncertainty leads us to a
discussion of systematics in the sense of training-related
uncertainties which do not shrink with more training data. Our
regression task naturally leads us to developing a framework to
calibrate deep network taggers and account for uncertainties in the
training sample. We find that a straightforward treatment should be
based on smearing the momentum labels in the training sample. It
directly accounts for the uncertainties in the underlying measurements
of the calibration sample and treats them as an additional systematic
effect on the top momentum measurement. As before, the Bayesian
network allows us to cleanly separate all different sources of
uncertainty.

Our simple application serves as an example how we can use Bayesian
networks to define statistical and systematic uncertainties coming
from the training sample and affecting the network output. These error
bars are defined jet by jet, or event by event, giving us more control
than standard methods do. Training on smeared labels allows us to
implement energy calibration in a straightforward and automized
manner. While our modelling of uncertainties on the reference
measurements for calibration is simplified, our approach can be
extended in a straightforward manner. For instance, the effect of
different jet algorithms or different Monte Carlo simulations can be
implemented as a non-Gaussian contribution to the label smearing. The
key observation is that Bayesian networks allow us to quote
uncertainties from all kinds of statistical and systematic limitations
of the labelled training data.

\section{Bayesian regression}
\label{sec:bayes}

While standard neural networks adapt a set of weights $\omega$ to
describe a general function based on some kind of training, Bayesian
networks learn weight
distributions~\cite{bnn_early,bnn_early2,bnn_early3,bnn_tev,bnn_tev2,bnn_nu}. Sampling
over those $\omega$-distributions gives us access to uncertainties in
the network output, induced by limitations of the training data. After
studying the effect of limited training statistics on jet
classification~\cite{ours}, we now generalize our approach to include
limited training statistics as well as the systematic effects from
stochastic or smeared training data.

As an example, we want to extract the transverse momentum $p_T$ of a
hadronically decaying top quark from a
fat top jet. 
%
If we define $p(p_T | j)$ as the probability over possible $p_T$ values for a given top jet, $j$, we can  extract the mean value as:
\begin{align}
\langle p_T \rangle 
= \int dp_T \; p_T \; p(p_T|j) \; .
\label{eq:folding}
\end{align}
For a Bayesian network $p(p_T|j)$ is generated by sampling over the
trained weight distributions $p(\omega | M)$,
\begin{align}
  p(p_T|j) 
  =
  \int d \omega \; p(p_T | \omega, j) \; p(\omega | M) \; ,
\label{eq:folding}
\end{align}
where $M$ is the training data set.
Obviously, we do not know the closed form of $p(\omega | M)$. In the
sense of a distribution~\cite{blei}, the network training will
approximate it with the learned function $q(\omega)$,
\begin{align}
p(p_T | j)
= \int d \omega \; p(p_T | \omega, j) \; p(\omega | M )
\approx \int d \omega \; p(p_T | \omega, j) \; q(\omega) \; .
\label{eq:first_bayes}
\end{align}
If we exchange the two integrals, the mean transverse momentum becomes
\begin{align}
\langle p_T \rangle 
&\equiv \int d\omega \; q(\omega) \langle p_T \rangle_\omega \
\quad \text{with} \quad
\langle p_T \rangle_\omega 
= \int dp_T \; p_T \; p(p_T | \omega, j) \; .
\label{eq:expecations}
\end{align}
%
Correspondingly, the variance of the $p_T$ extraction can be extracted as
\begin{align}
\sigma_\text{tot}^2
&= \langle \left( p_T - \langle p_T \rangle \right)^2 \rangle \notag \\
&= \int d\omega \; q(\omega) \left[ 
   \langle p_T^2 \rangle_\omega 
 - 2 \langle p_T \rangle \langle p_T \rangle_\omega
 + \langle p_T \rangle^2 \right] \notag \\
&= \int d\omega \; q(\omega) \left[ 
   \langle p_T^2 \rangle_\omega - \langle p_T \rangle_\omega^2
 + \left( \langle p_T \rangle_\omega - \langle p_T \rangle \right)^2 \right] 
\equiv \sigma_\text{stoch}^2 + \sigma_\text{pred}^2 \; .
\label{eq:def_sigmas}
\end{align}
This is the critical step which allows us to identify two
contributions to the jet-wise uncertainty from the Bayesian
network. First, a finite $\sigma_\text{stoch}$ occurs without even
sampling the network weights, so it describes a systematic effect from
the stochastic nature of the training sample,
\begin{align}
\sigma_\text{stoch}^2 
\equiv \langle \sigma_{\text{stoch}, \omega}^2 \rangle 
=& \int d\omega \; q(\omega) \; \sigma_{\text{stoch}, \omega}^2 \notag \\
=& \int d\omega \; q(\omega) \left[ 
   \langle p_T^2 \rangle_\omega - \langle p_T \rangle_\omega^2 \right] \; .
\label{eq:sig_stoch}
\end{align}
Second, $\sigma_\text{pred}$ is defined in terms of the
$\omega$-integrated expectation value $\langle p_T \rangle$, so there
does not exist an $\omega$-dependent version,
\begin{align}
\sigma_\text{pred}^2
&= \int d\omega \; q(\omega) 
   \left( \langle p_T \rangle_\omega - \langle p_T \rangle \right)^2 \; .
\label{eq:sig_pred}
\end{align}
Only this second contribution will vanish in the limit of an
infinitely large training sample, because in that case the network
weight distributions become delta distributions. We will discuss the
nature of these two contributions in detail in our analysis.

The two contributions to $\sigma_\text{tot}$ can also be identified in
the loss function. The standard approach for Bayesian networks is to
start with Eq.\eqref{eq:first_bayes} implemented as a Kullback-Leibler
divergence,
\begin{align}
\kl [q(\omega),p(\omega|M)] 
&= \int d\omega \; q(\omega) \; \log \frac{q(\omega)}{p(\omega|M)} \notag \\
&= \int d\omega \; q(\omega) \; \log \frac{q(\omega) p(M)}{p( M|\omega) p(\omega)} \notag \\
&= \underbrace{\kl[q(\omega),p(\omega)] 
 - \int d\omega \; q(\omega) \; \log p(M|\omega)}_{\equiv L_\kl}
 + \log p(M) \int d\omega \; q(\omega) \; .
\label{eq:kl_bayes}
\end{align}
%
In this derivation we use Bayes' theorem. The prior $p(\omega)$
describes the model parameters before training. The model evidence
$p(M)$ guarantees the correct normalization of $p(\omega | M )$.
Turning Eq.\eqref{eq:kl_bayes} into a loss function we can omit it
just as the normalization condition for $q(\omega)$. The relevant loss
function of the Bayesian network, $L_\kl$, then consists of two terms,
the regularization for $q(\omega)$ in reference to the prior
$p(\omega)$ and the likelihood $p(M|\omega)$, which we can work with
in a frequentist sense. For a Gaussian prior the regularization term
becomes the standard L2-regularization.

\begin{figure}[t]
\centering
\includegraphics[width=0.80\textwidth]{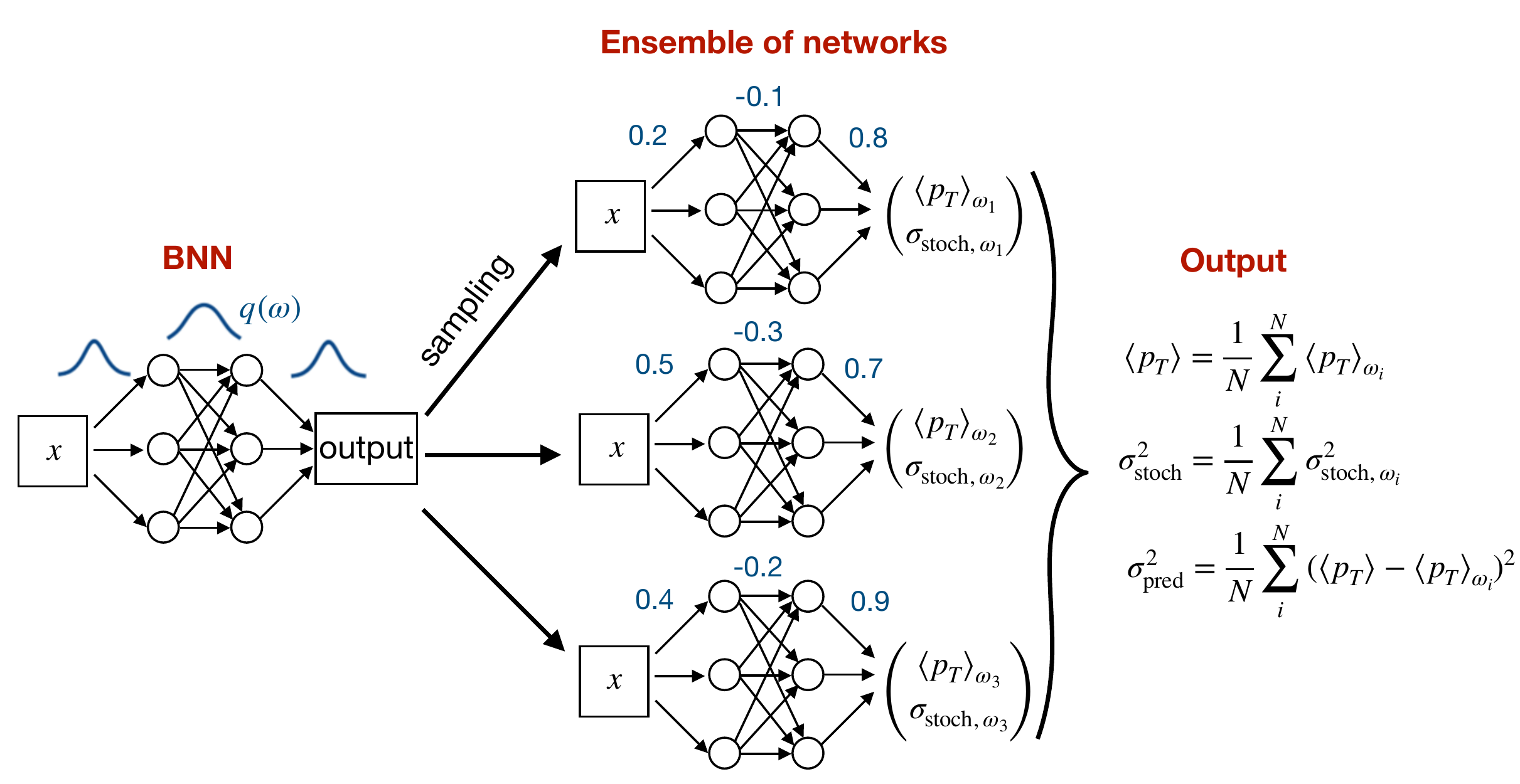}
\caption{Illustration of our Bayesian network setup. The Bayesian
  network provides us with an uncertainty estimate for a single input
  jet $x$.}
\label{fig:bnn}
\end{figure}

For illustration purposes or to improve the numerical performance we
can now make a set of assumptions. In Ref.~\cite{ours} we have shown,
by varying priors over several order of magnitude, that assuming a
Gaussian prior $p(\omega)$ had no visible effect on the network
output. To get analytic control, we can approximate the likelihood
$p(M|\omega)$ as Gaussian,
\begin{align}
 \log p(M|\omega) 
\approx - \frac{\left(p_{T}^{\mathrm{truth}} - \langle p_{T} \rangle_{\omega}\right)^2}{2\sigma_{\text{stoch}, \omega}^2} - \frac{1}{2} \log \sigma_{\text{stoch}, \omega}^2 + \text{const} \; .
\label{eq:neg_log}
\end{align}
where $p_{T}^{\mathrm{truth}}$ is the truth label provided by the training data set $M$. The width of this Gaussian corresponds to a systematic uncertainty, so
we identify it with $\sigma_{\text{stoch},\omega}$.  The loss function
\begin{align}
L_\kl
\approx \kl[q(\omega),p(\omega)] 
 + \int d\omega \; q(\omega) \; 
   \left[  \frac{\left( p_{T}^{\mathrm{truth}} -  \langle p_{T} \rangle_{\omega} \right)^2}{2\sigma_{\text{stoch}, \omega}^2} + \frac{1}{2} \log \sigma_{\text{stoch}, \omega}^2 
   \right] \; ,
\label{eq:kl_loss1}
\end{align}
now has to be minimized with respect to the parameters of
$q(\omega)$. Because we have assumed $q(\omega)$ to be Gaussian that
gives us two trainable parameters per weight, and our neural
network gives
\begin{align}
\text{NN} (\omega) = 
\begin{pmatrix}
\langle p_T \rangle_{\omega}\\
\sigma_{\text{stoch}, \omega}
\end{pmatrix} 
\label{eq:output}
\end{align}
per jet.
To extract the per-jet probability distribution $p(p_T| x)$ following
Eq.\eqref{eq:first_bayes}, we usually rely on Monte Carlo integration
by sampling weights from the weight distributions.  As in
Eq.\eqref{eq:kl_loss1} we assume that $p(p_T| \omega, x)$ is a
Gaussian with the above-defined mean $\langle p_T \rangle_\omega$ and
width $\sigma_{\text{stoch}, \omega}$.  Moreover, for large training
statistics the distribution $q(\omega)$ should become narrow.
According to Eq.\eqref{eq:sig_pred} the effect of a finite width of
$q(\omega)$ can be tracked by $\sigma_\text{pred}$, so in the limit
$\sigma_\text{pred} \ll \sigma_\text{stoch}$ we can approximate
$p(p_T| x)$ as a Gaussian with weight-independent mean $\langle p_T
\rangle$ and width $\sigma_\text{stoch}$.  This network structure is
illustrated in Fig.~\ref{fig:bnn}.

\section{Data set and network}
\label{sec:data_net}

The correct and precise reconstruction of the momentum of tagged top
quarks is important for instance in top resonance searches and has
influenced the design of many top taggers~\cite{heptop}. Our data set
is therefore similar to standard top tagging references, with some
modifications which simplify our regression task.  We generate a
sample of $R=1.2$ top jets in the range $\ptt = 400~...~1000$~GeV with
\textsc{Pythia}~\cite{pythia} at 14~TeV collider energy and the
standard ATLAS card for \textsc{Delphes}~\cite{delphes}.  We always
neglect multi-parton interactions and always include final state
radiation. Given initial state radiation we work with two event
samples, one with ISR switched on and one with ISR switched off.  We
require the jets to be central $|\eta_j| < 2$ and truth-matched in the
sense that each fat jet has to have a top quark within the jet
area. These settings essentially correspond to the public top tagging
data set from Refs.~\cite{lola} and~\cite{jets_comparison}.  The
difference to the standard tagging reference sets is that we flatten
our data set in $\ptt$, such that even accounting for bin migration
effects we can safely assume that in the fat jet momentum the sample
is flat for $p_{T,j} = 500~...~800$~GeV. 

The final result of our Bayesian network will be a probability
distribution over possible $p_{T,t}$ values for a given jet. For our
labelled data we know the corresponding $\ptt$.  However, the fact
that we will modify this truth label as part of the calibration
training makes it the less attractive option to organize our
samples. The closest alternative observable is the momentum of the fat
jet, so we can think of $p_{T,j}$ as representing the complete fat jet
input to the network. So unless explicitly mentioned we train our
networks on a large data set defined in terms of the fat jet momentum,
\begin{align}
p_{T,j} = 400~...~1000~\gev \qquad \text{(training sample)} \; ,
\end{align}
Whenever we need a homogeneous sample without boundary effects we
choose a narrow test sample with
\begin{align}
p_{T,j} = 600~...~620~\gev \qquad \text{(narrow test sample)} \; 
\end{align}

The data format for the fat jet information is a $p_T$-ordered list of
up to 200 constituent 4-vectors ($\vec{p}$ and $E$) with ISR and 100
constituents without. Our total sample size is 2.2M jets without ISR,
of which we use 400k jet for validation and testing, each. The
training size is varied throughout our analysis.\

Our regression network is a simple 5-layer fully connected dense
network. Its first two layers each consist of 100 units, the next two
50 units, followed by a 2-unit output layer, unless mentioned
otherwise. For the prior we choose a Gaussian around zero and with
width 0.1. We have confirmed that our results are width-independent
over a wide range~\cite{ours}.  The typical sizes and widths of the
weights depends on the input data. The input is a flattened set of
4-vectors where we re-scale the $p_T$ values by a factor 1000 to end
up between zero and one. The activation function is ReLU, except for
the output layer. That one predicts the mean value $\langle p_T
\rangle$ without any need for an activation function and the SoftPlus
function for the error to have a smooth function which guarantees
positive values for the error. We have checked that this setup with
these hyper-parameters is not fine-tuned.

For the Bayesian network features we rely on Tensorflow
Probability~\cite{tensorflow_probability} with Flipout Dense
layers~\cite{flipout} replacing the dense layer of the deterministic
network.  All networks are trained with the Adam optimizer~\cite{adam}
and a learning rate of $10^{-4}$, determined by early stopping when
the loss function evaluated on the training dataset does not improve
for a certain number of epochs. This patience was set to 10 for a
training size of 1M jets and to larger values for smaller training
sizes because the loss function is more fluctuating. For the Bayesian
network with a training batch size of 100 we observe no over-fitting.

\section{Momentum determination and statistics}
\label{sec:stat}

As a first part of our Bayesian regression analysis we need to show
how well the networks reconstructs the top momentum and what the
limiting factors are. We then have to separate the statistical and
systematic uncertainties. In analogy to Ref.~\cite{ours} we first
study how the size of the training sample affects the regression
output, \ie how well the Bayesian network keeps track of the
statistical uncertainty.

\begin{figure}[t]
\centering
\includegraphics[width=0.48\textwidth]{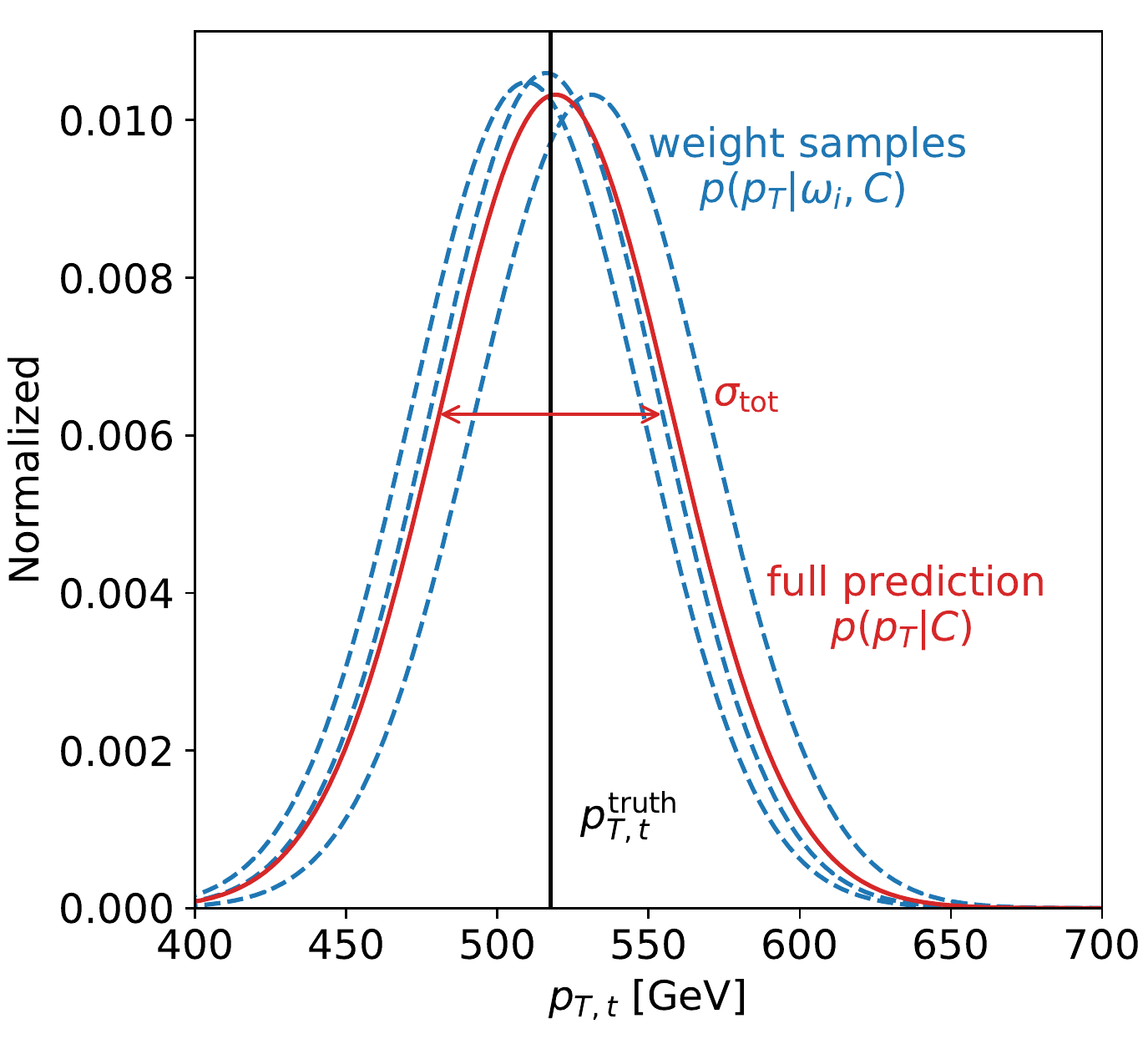}
\caption{Illustration of the predicted distribution from our Bayesian
  setup for a single top jet. We show the individual predictions from
  sampling the weights (petrol) as well as the aggregate prediction
  (red) and the corresponding per-jet uncertainty
  $\sigma_\text{tot}$.}
\label{fig:illustration}
\end{figure}

To illustrate the output of our Bayesian network for a single jet we
shoe an example in Fig.~\ref{fig:illustration}.  Sampling from the
weight distributions, $q(\omega)$, provides us with a Gaussian per
sampled set of weights, shown in petrol. The combination of these
distributions is shown in red.  The width of the combined distribution
is the predicted per-jet uncertainty $\sigma_\text{tot}$, defined in
Eq.\eqref{eq:def_sigmas}. For illustration purposes we pick a top jet
where $\ptt$ coincides with the peak of the predicted distribution.

\subsubsection*{Regression performance}

\begin{figure}[t]
\centering
\includegraphics[width=0.48\textwidth]{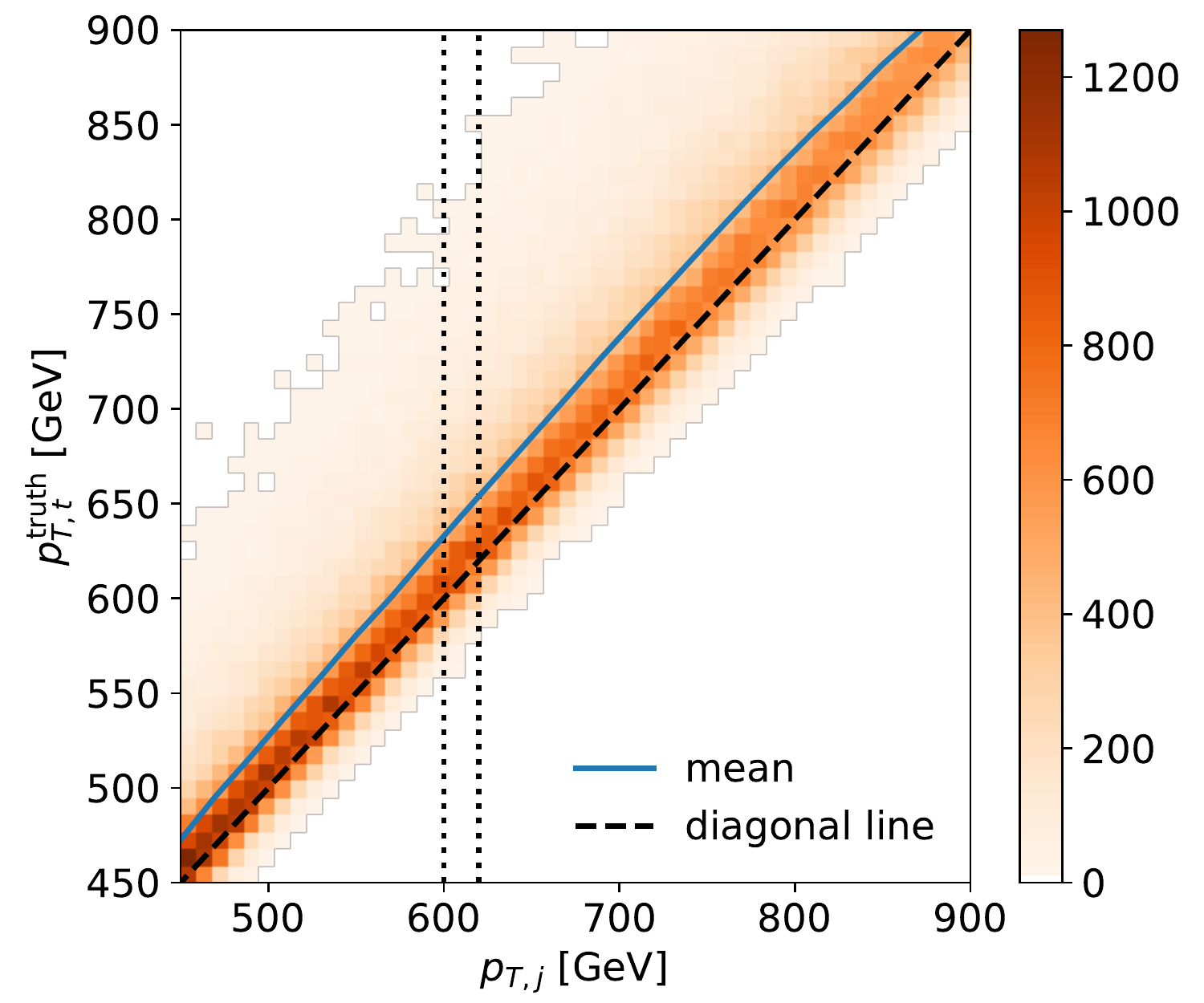}
\hspace*{0.01\textwidth}
\includegraphics[width=0.48\textwidth]{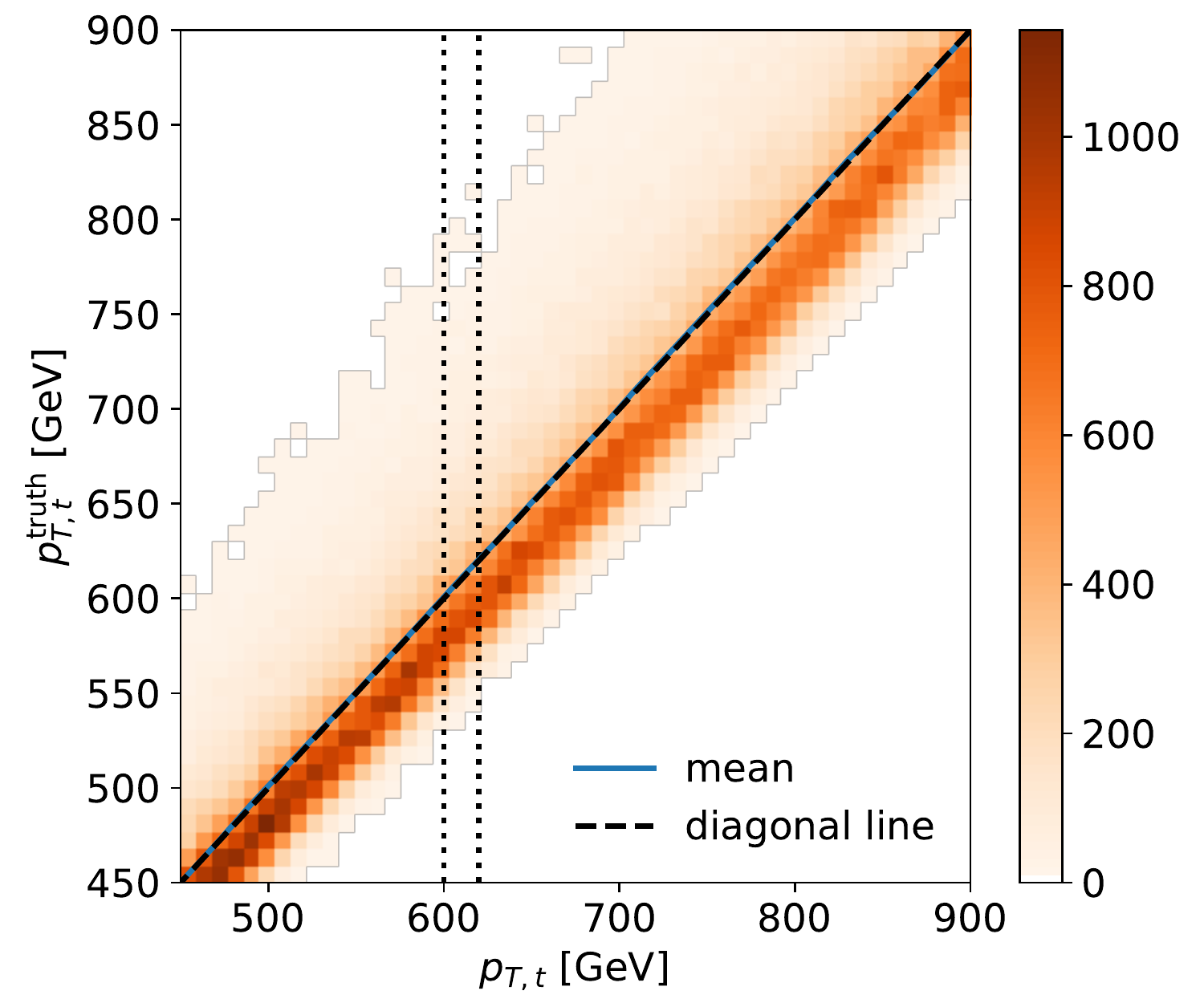}
\caption{Correlation between the fat jet's $p_{T,j}$ and the truth
  label $\ptt$ (left) and between the extracted $p_{T,t}$ and the
  truth label $\ptt$ (right). Both correlations are shown with initial
  state radiation in the training and test samples switched off.}
\label{fig:pt2pt}
\end{figure}

To begin, we show in the left panel of Fig.~\ref{fig:pt2pt} the
correlation between the measurable $p_{T,j}$ and the MC label
$\ptt$. We see that over the entire range the two values are aligned
well. This allows us to use $p_{T,j}$ as a proxy to the truth
information, keeping in mind that we will eventually smear the truth
label to describe the jet calibration.  In the right panel of
Fig.~\ref{fig:pt2pt} we show the correlation between the central
extracted $p_{T,t}$ value, which in Sec.~\ref{sec:bayes} is properly
denoted as the expectation value $\langle p_T \rangle$, and the label
$\ptt$. 

\begin{figure}[t]
\centering
\includegraphics[width=0.48\textwidth]{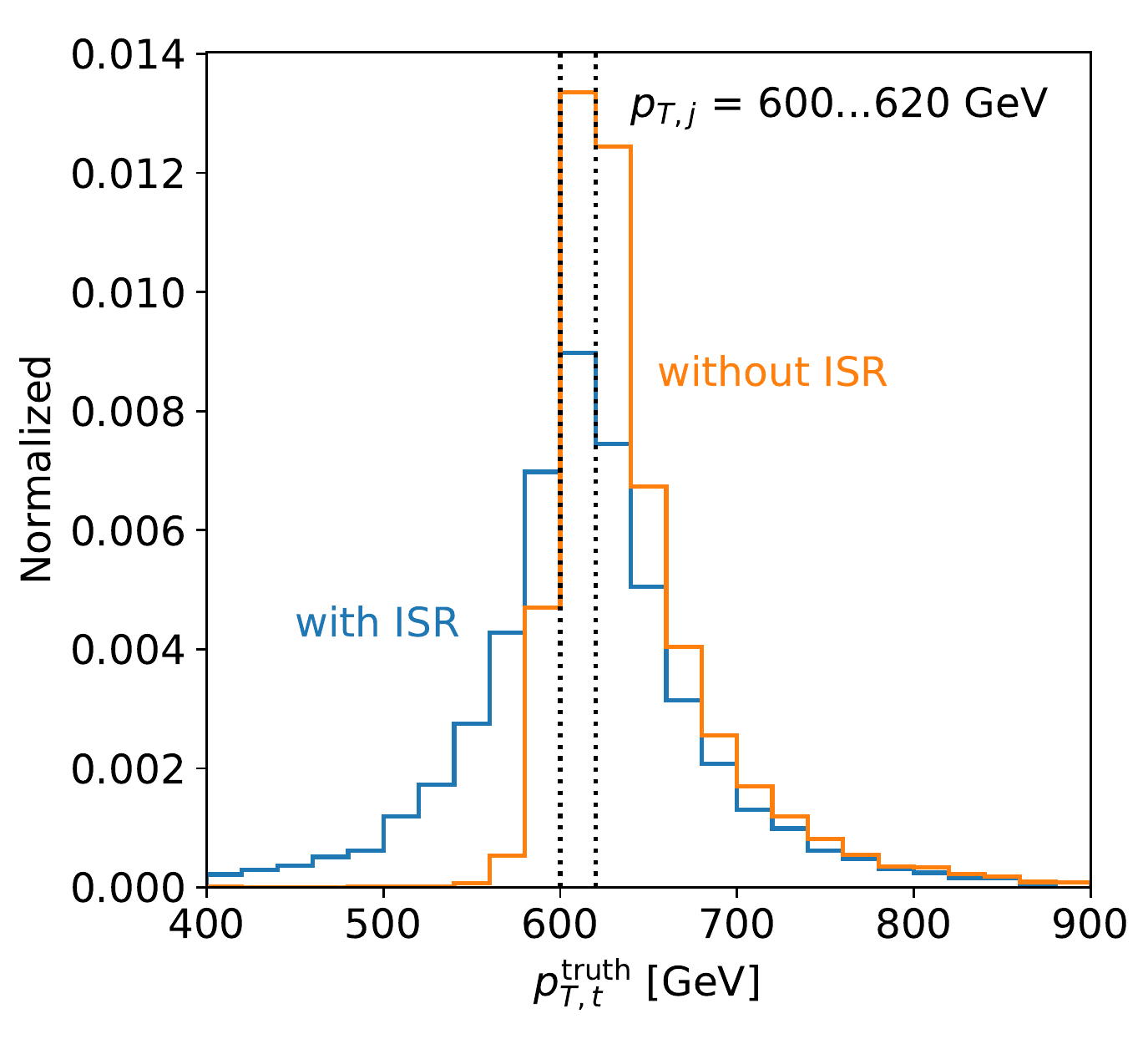}
\hspace*{0.01\textwidth}
\includegraphics[width=0.48\textwidth]{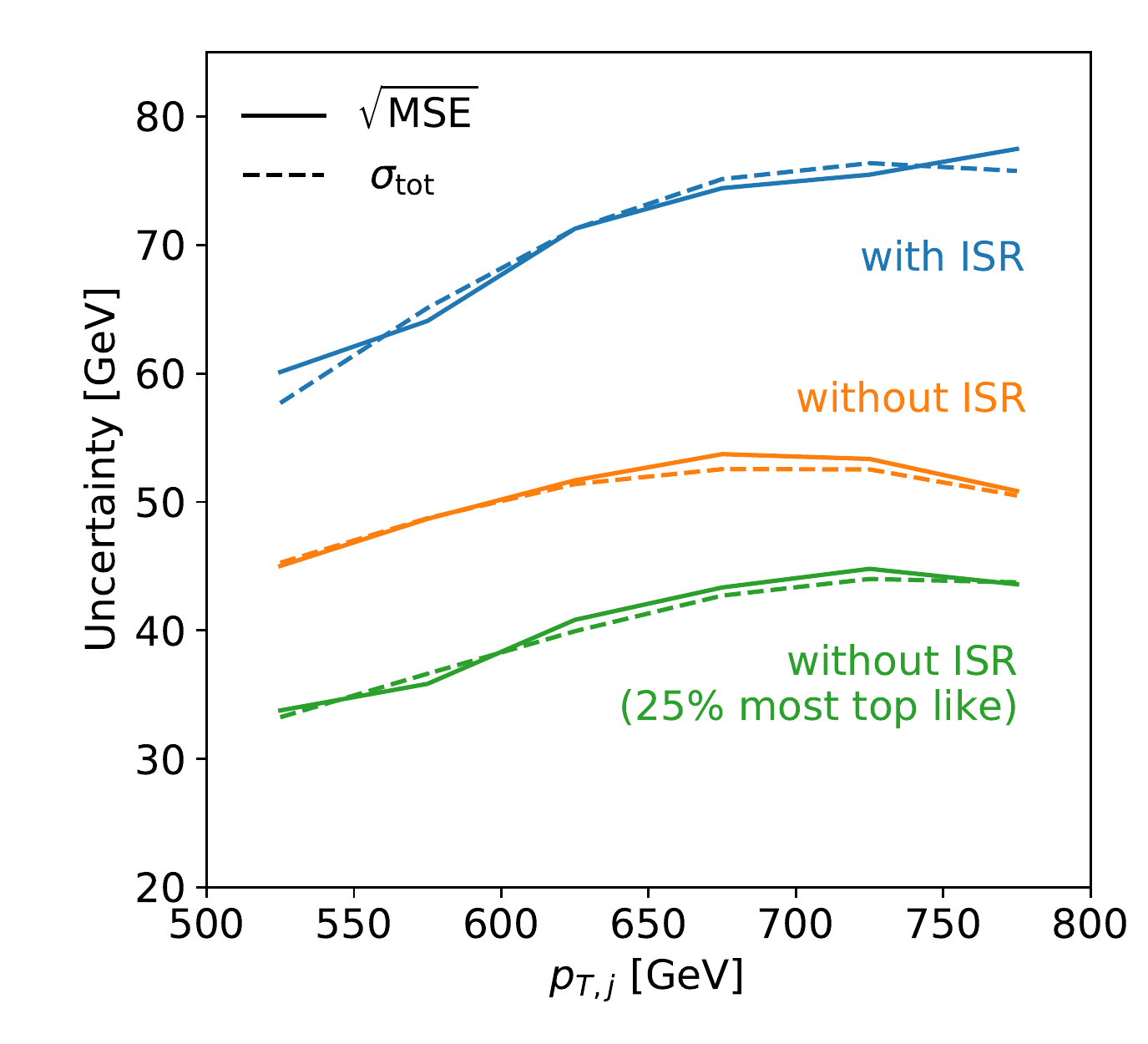}
\caption{Left: distribution of the truth label $\ptt$ for jets with
  $p_{T,j} = 600~...~620$~GeV, without and with initial state
  radiation.  Right: regression uncertainty as a function of $p_{T,j}$
  (solid), compared with the average $\sigma_\text{stoch}$ as the
  network output (dashed). The most top-like events are defined with a
  simple \text{LoLa} tagger~\cite{lola}.}
\label{fig:performance}
\end{figure}

In the left panel of Fig.~\ref{fig:performance} we show the $\ptt$
distribution for the narrow slice $p_{T,j} = 600~...~620$~GeV. In the
absence of initial state radiation the distribution is asymmetric. The
simple reason is that the jet clustering can only miss top decay
constituents, so we are more likely to observe $p_{T,j} < \ptt$. Aside
from that we see a clear peak, suggesting that we can indeed represent
$\ptt$ with $p_{T,j}$.  Because the peak is washed out by ISR, we
switch off ISR to make it easier to understand the physics behind our
network task. In practice, this could be done through a pre-processing
and grooming step.

\begin{table}[b!]
\centering
\begin{small} \begin{tabular}{c|cc|cc}
\toprule
$p_{T,j}=600~...~620$~GeV  
& $\mse$ & $\mse/p_{T,j}$  & $\mse$ & $\mse/p_{T,j}$ \\
& \multicolumn{2}{c|}{With ISR} & \multicolumn{2}{c}{Without ISR} \\ 
\midrule 
All jets           & $69.7\pm0.2$  & $(11.43\pm 0.03)\%$ 
                   & $50.6\pm0.1$  & $(8.30\pm 0.02)\%$ \\ 
75\% most top-like & $67.8\pm 0.2$ & $(11.11\pm 0.01)\%$ 
                   & $45.5\pm 0.1$ & $(7.47\pm 0.02)\%$ \\ 
50\% most top-like & $66.5\pm 0.1$ & $(10.89\pm 0.01)\%$ 
                   & $41.8\pm 0.1$ & $(6.85\pm 0.01)\%$ \\  
25\% most top-like & $66.5\pm 0.1$ & $(10.89\pm 0.02)\%$ 
                   & $40.4\pm 0.1$ & $(6.63\pm 0.02)\%$ \\  
\bottomrule
\end{tabular} \end{small}
\caption{Performance of $p_{T,t}$ regression, uncertainty representing
  the standard deviation of 5 trainings. The narrow $p_{T,j}$ range
  refers to the 5k test jets, not the 500k training jets.}
\label{tab:performance}
\end{table}

Whenever we have access to MC truth, we can measure the performance of
the regression network for each top jet as $(p_{T,t} - \ptt)^2$.  The
squared difference measure only uses the mean or central value
reported by a Bayesian or deterministic network, not the additional
uncertainty information from the Bayesian network. For a given test
sample with $N$ top jets $t_i$ we construct the mean quadratic error
as
\begin{align}
\mse = 
\left[ \frac{1}{N} \sum_\text{jets $i$} \left(
 p_{T,t_i} - p_{T,t_i}^\text{truth} 
\right)^2 \right]^{1/2} \,
\label{eq:mse}
\end{align}
We evaluate it over homogeneous samples, for example our usual slice
in $p_{T,j}$. In Tab.~\ref{tab:performance} we contrast results with
and without ISR and show what happens if we limit ourselves to the
most top-like jets based on a standard LoLa tagger~\cite{lola},
trained on events with ISR.  To estimate the effect of different
trainings we also give an error bar based on five independent
trainings and the resulting standard deviation. Expectedly, the
$p_T$-measurement benefits from more top-like events, but the effect
is not as significant as in the \textsc{HEPTopTagger}
analysis~\cite{heptop}. One of the reasons is that we are using
relatively large $R=1.2$ jets for the high transverse momentum
range. Similarly, we confirm that additional ISR jets have the
potential to affect the top momentum measurement whenever hard extra
jets enter the fat jet area.

In the right panel of Fig.~\ref{fig:performance} we show $\mse$ as a
function of $p_{T,j}$ for a bin width of 40~GeV. While the absolute
error increases, the relative error on the extracted $p_{T,t}$ shrinks
for more boosted jets. If we assume that an improved jet pre-selection
can efficiently remove ISR contributions our regression network can
measure the top momentum to roughly 4\%. This result is only a rough
benchmark to confirm that the regression network performs in a
meaningful manner. It would surely be possible to improve the network
performance, but we deliberately keep the network simple, to
understand the way it processes information and the related
uncertainties.  From the right panel of Fig.~\ref{fig:performance} we
know that boundary effects will appear already around 200~GeV away
from the actual boundaries. Indeed, around $p_{T,j}$ we see such
effects indicating the phase space boundary of $p_{T,j} < 1$~TeV in
our training sample.

In the same Fig.~\ref{fig:performance} we also show this uncertainty
estimate of the Bayesian network, $\sigma_\text{tot}$ as defined in
Eq.\eqref{eq:def_sigmas}. It follows the $\mse$ estimate of the
network error, indicating that the Bayesian output captures the same
physics as the frequentist-defined spread of the central values.

\subsubsection*{Training sample size and $\sigma_\text{pred}$}

\begin{figure}[t]
\centering
\includegraphics[width=0.48\textwidth]{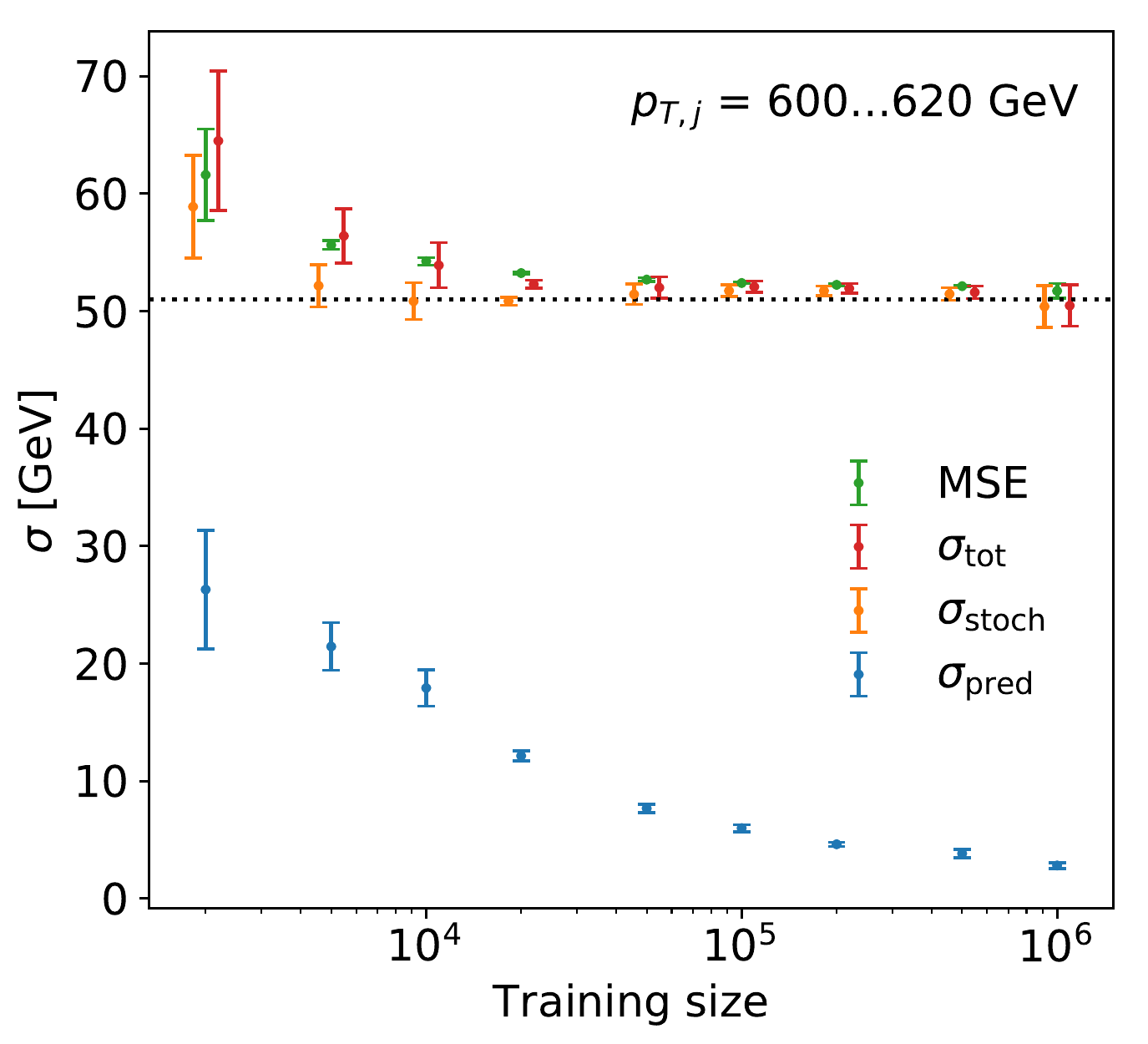}
\caption{Uncertainty contributions $\sigma_\text{pred}$ and
  $\sigma_\text{stoch}$ as a function of the size of the training
  sample. The error bar represents the standard deviation of five
  different trainings. In addition we include $\mse$ as defined in
  Eq.\eqref{eq:mse}.}
\label{fig:training_size}
\end{figure}

As discussed in Sec.~\ref{sec:bayes} the contribution
$\sigma_\text{pred}$ to the uncertainty reported by the network can be
identified as a statistical uncertainty in the sense that it should
vanish in the limit of infinitely many training jets. In complete
analogy to the classification task described in Ref.~\cite{ours} we
confirm this by training Bayesian networks on 2k, 5k, 10k, 15k, 20k,
30k, 50k, 100k, 200k, 500k, and 1M jets. We test these networks on the
narrow range $p_{T,j} = 600~...~620$~GeV, similar to the results shown
in Tab.~\ref{tab:performance}. The uncertainties quoted by the
Bayesian network are shown in Fig.~\ref{fig:training_size}. In the
lower part of the figure we first see that the statistical error
$\sigma_\text{pred}$ indeed asymptotically approaches zero for
1M training jets.  The error bars on the extracted uncertainty are
given by the standard deviation of five independent trainings.  As
expected, they grow for smaller training samples, where the Bayesian
networks also give fluctuating results.

In the same figure we also show the systematic $\sigma_\text{stoch}$
and the combined $\sigma_\text{tot}$, defined in
Eq.\eqref{eq:def_sigmas}. We confirm that the extracted
$\sigma_\text{stoch}$ hardly depends on the size of the training
sample. Once we have a reasonably number of training events it reaches
a plateau of around 50~GeV or 8\%, while for less than 10000 training
events the network simply fails to capture the full information. We can
compare the plateau value for $\sigma_\text{stoch}$ to the $\mse$
value and find again that the two values agree.This allows us to
conclude that $\sigma_\text{stoch}$ describes a systematic
uncertainty and that it is related to the truth-based $\mse$
estimate. We will discuss it in more detail in Sec.~\ref{sec:syst}.

\begin{figure}[t]
\centering
\includegraphics[width=0.325\textwidth]{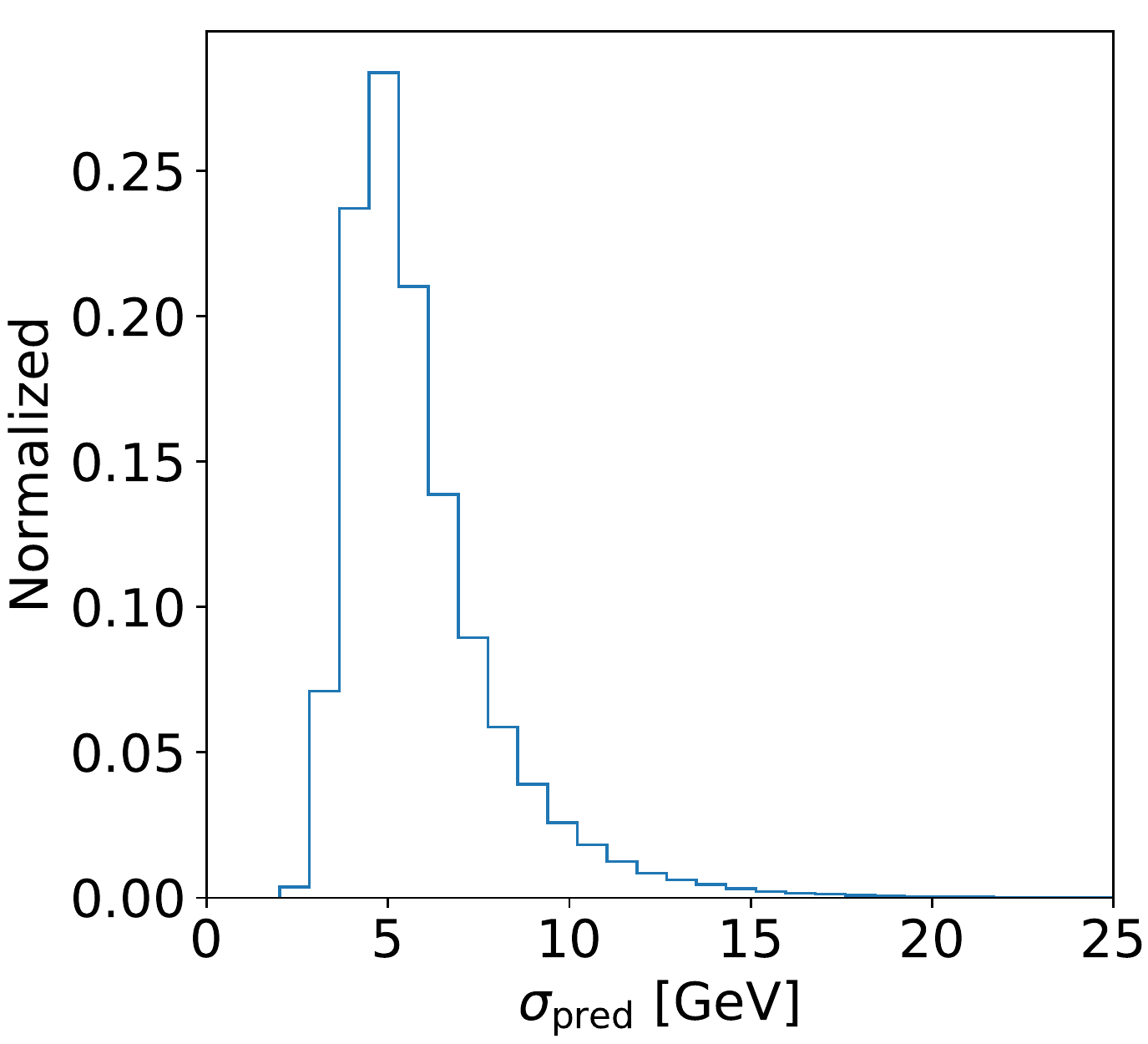}
\includegraphics[width=0.325\textwidth]{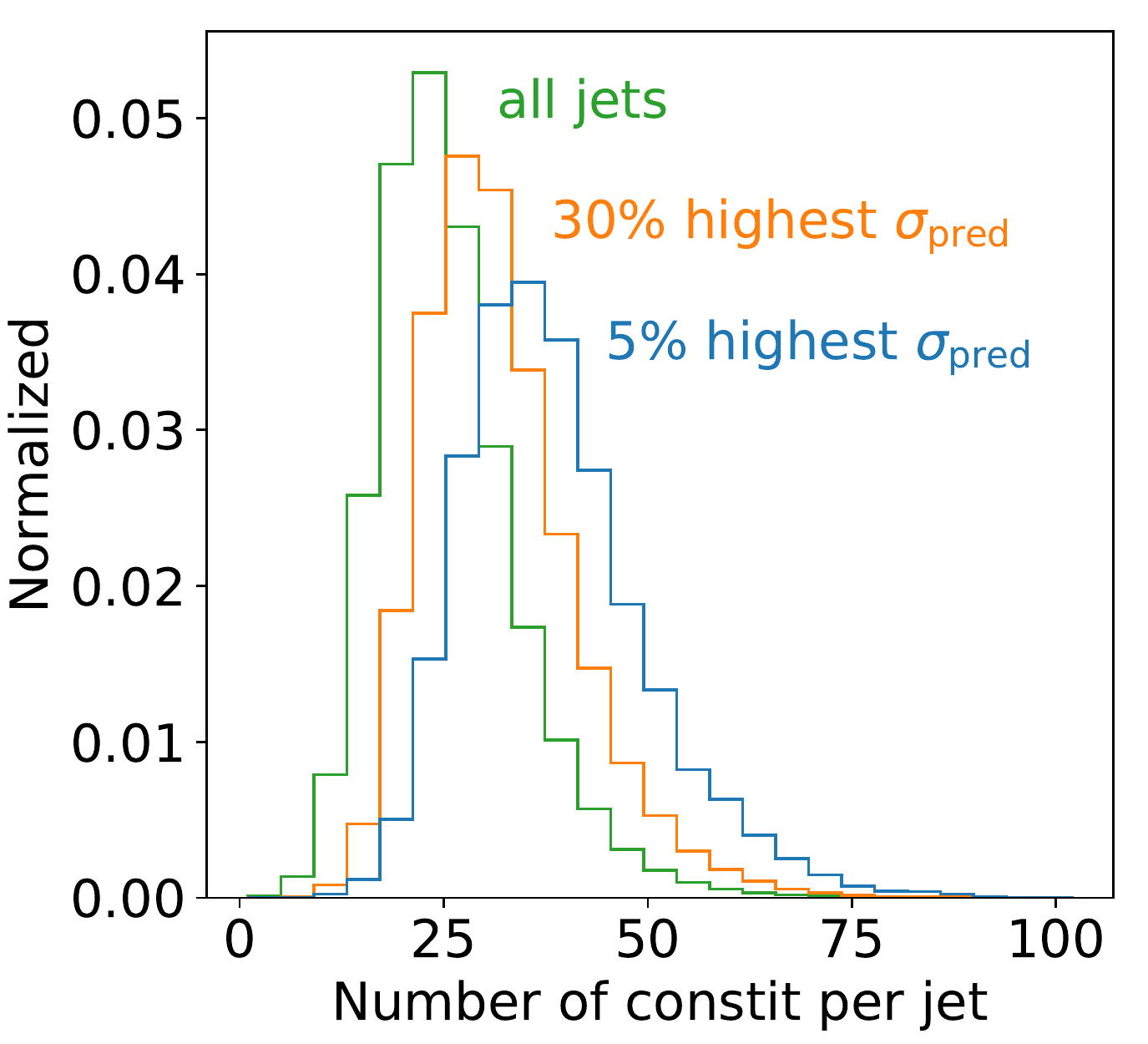}
\includegraphics[width=0.325\textwidth]{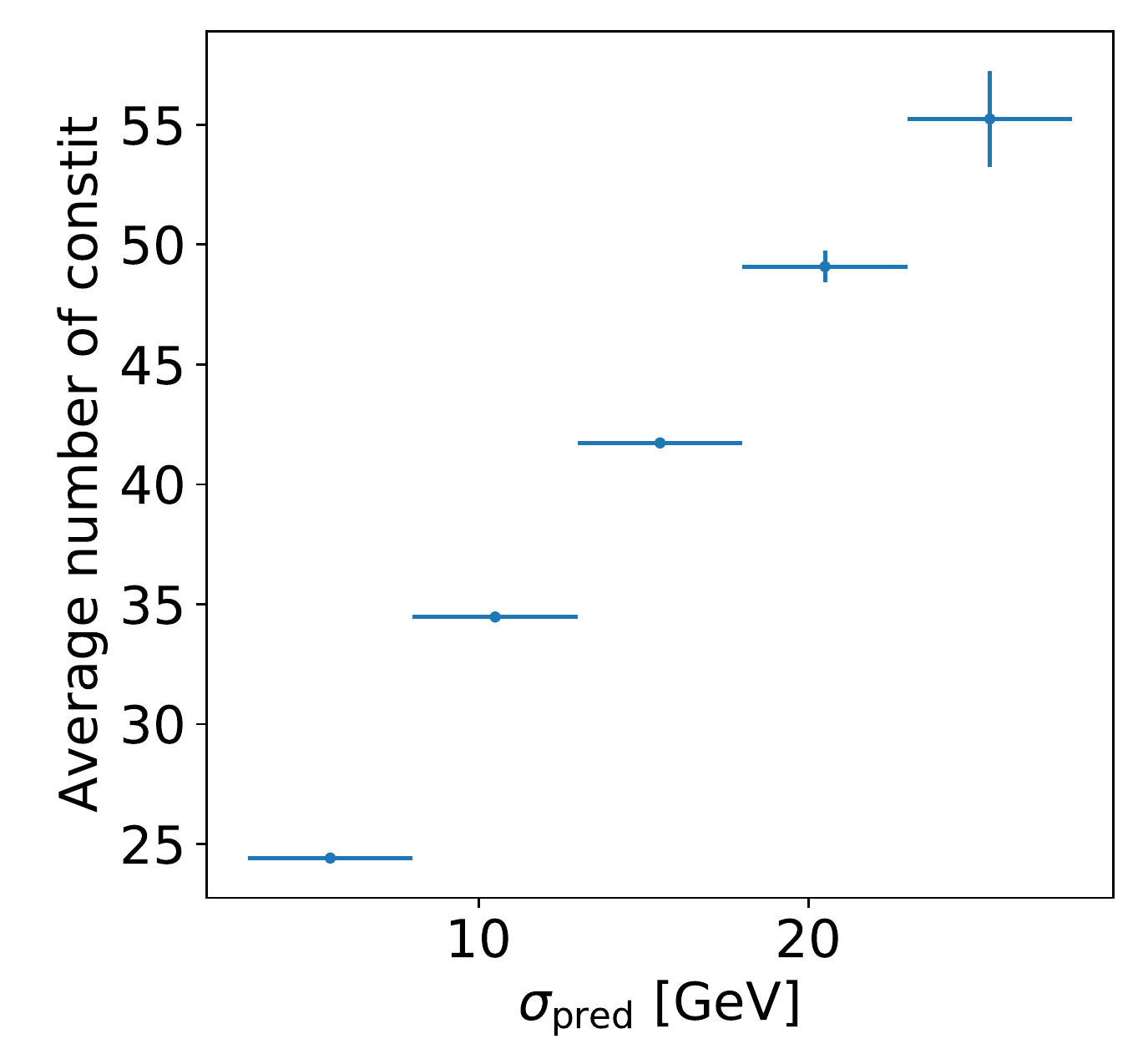}
\caption{Left: distribution of the statistical uncertainty
  $\sigma_\text{pred}$ for 400k jets. Center: number of constituents
  per jet for different $\sigma_\text{pred}$. Right: average number of
  constituents per jet as a function of the extracted statistical
  uncertainty.}
\label{fig:n_constit_simga_pred}
\end{figure}

After observing the average effect of the training sample size on
$\sigma_\text{pred}$ the obvious question is if we can understand this
behavior. In the left panel of Fig.~\ref{fig:n_constit_simga_pred} we
show the distribution of $\sigma_\text{pred}$ values for a sample of
400k jets. The network is trained on 100k jets with an extended range
$p_{T,j} = 500~...~900$~GeV. We see a clear maximum around
$\sigma_\text{pred} \approx 5$~GeV, with a large tail towards large
uncertainties. It is induced by the constraint that no network should
quote an uncertainty close to zero.

The jet property we can relate to the $\sigma_\text{pred}$ behavior is
the number of particle-flow constituents. As mentioned before, we
cover up to 100 constituents for jets without ISR. Their effect on top
tagging is discussed for instance in Ref.~\cite{lola}. The center
panel of Fig.~\ref{fig:n_constit_simga_pred} shows how the number of
constituents in the test sample jets peaks at around 25, but with a
tail extending to 60. Jets with a larger quoted uncertainty have
significantly more constituents. The same information is shown in the
right panel, where we see the average number of jets increases with
the range of quoted statistical uncertainties. The reason for this
pattern is that also within the training sample the number of
constituents will peak around 25, limiting the number of training jets
with higher constituent numbers. We note that we could use the same
argument using the jet mass.

\subsubsection*{Frequentist approach}

\begin{figure}[t]
\centering
\includegraphics[scale=0.48]{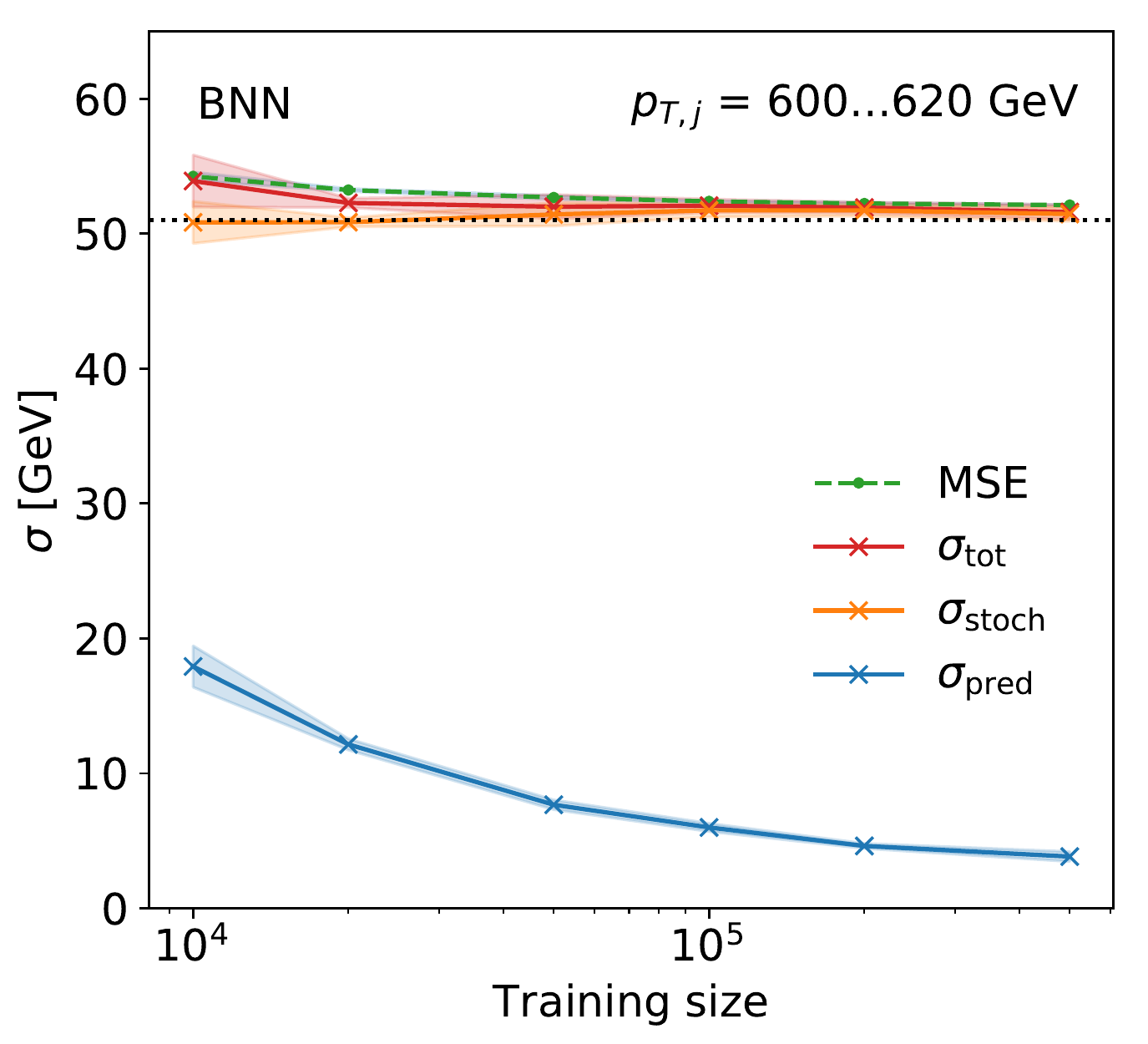}
\hspace*{0.01\textwidth}
\includegraphics[scale=0.48]{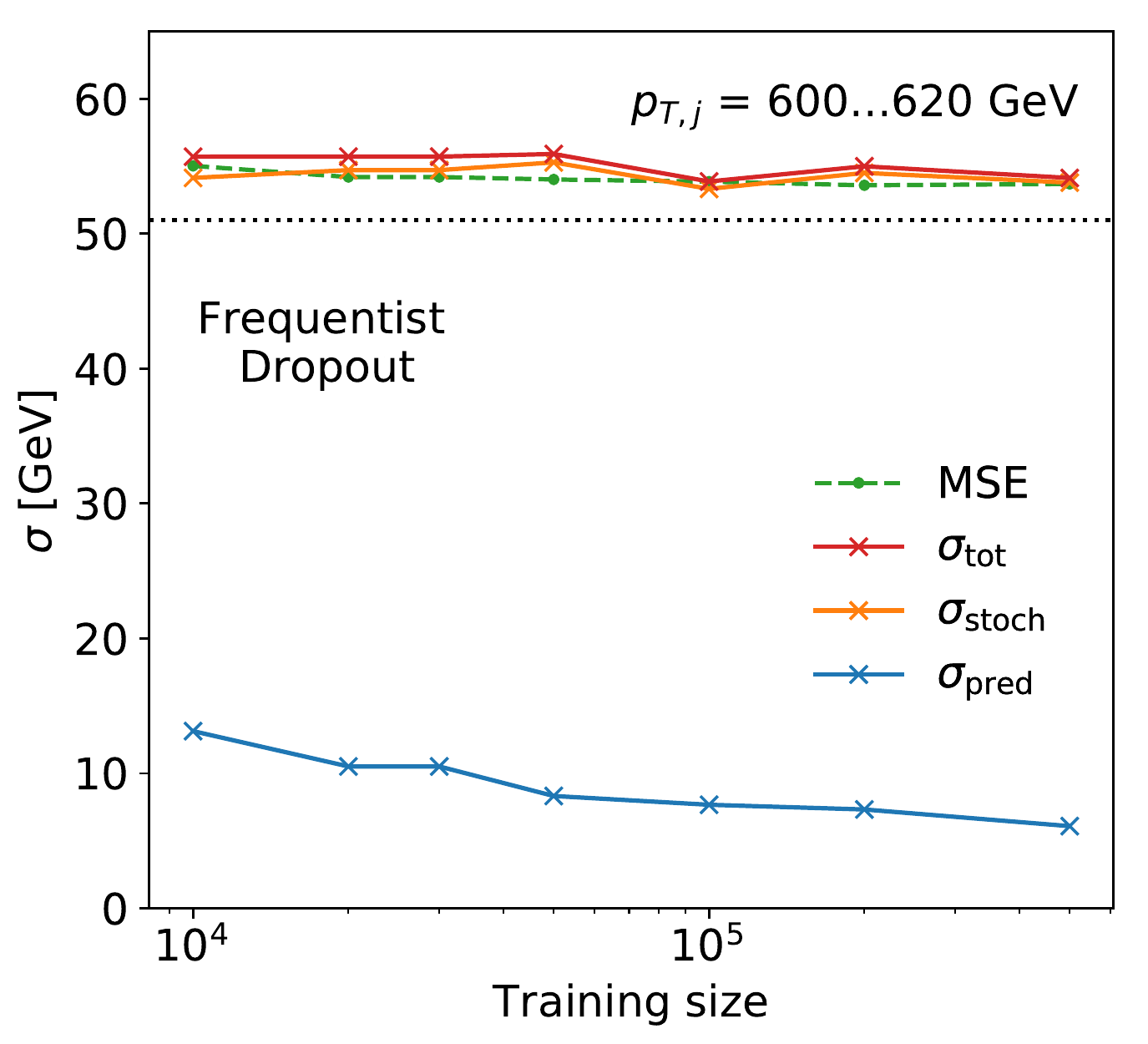}
\caption{All uncertainties as a function of the training size,
  comparing the Bayesian network (left) with a (frequentist) set of
  deterministic networks (right).  The left panel corresponds to
  Fig.~\ref{fig:training_size}, and the ranges indicate the standard
  deviation for five trainings.}
\label{fig:comp_frequentist}
\end{figure}

From a practical point of view it is crucial to validate the Bayesian
network using a frequentist approach. We do this by showing that
predictions from many trainings of a deterministic network reproduce
our Bayesian network results for the statistical uncertainty
$\sigma_\text{pred}$.

For the deterministic networks we use the same architecture as for the
Bayesian network. The loss function of the deterministic networks is
the negative log-likelihood given in Eq.\eqref{eq:neg_log}, and we fix
the L2-regularization to match the Bayesian network in
Eq.\eqref{eq:kl_bayes},
\begin{align}
\lambda_\text{L2} = \frac{1}{2 \sigma_\text{prior} N} \; ,
\end{align}
where $N$ is the total training size and $\sigma_\text{prior} = 0.1$
is our prior width. We then train 40 deterministic networks on
statistically independent samples, which we sample from the total of
2.2M training jets.  Each set of deterministic network then predicts a
mean and a standard deviation, in analogy to Eq.\eqref{eq:output}. The
difference between the Bayesian evaluation and the frequentist
networks is that we replace the integral over weights with a sum over
independent networks.  

For deterministic networks we need to avoid over-training. An
over-trained set of networks will underestimate $\sigma_\text{stoch}$,
while the spread represented by $\sigma_\text{pred}$
increases. However, it is not guaranteed that these two effects
compensate each other for finite training time.  This is why we
introduce dropout for each inner layer with a rate of 0.1.  This value
is a compromise between network performance and over-training. Unlike
in our earlier study~\cite{ours} we do not use a MAP modification of
the Bayesian network.

In Fig.~\ref{fig:comp_frequentist} we compare the Bayesian and
frequentist uncertainties for different training sample size. While
the results agree well for properly trained networks or large training
samples, the frequentist approach slightly underestimates the
uncertainty for small training samples. The plateau value of
$\sigma_\text{stoch}$ depends on the chosen dropout value.  Accounting
for this effect we see that the training-size-dependent
$\sigma_\text{pred}$ and the plateau value of $\sigma_\text{stoch}$,
agree well between the Bayesian network and the frequentist sanity
check.

\section{Systematics and calibration}
\label{sec:syst}

In our original paper~\cite{ours} we have shown that the Bayesian
setup propagates uncertainties from statistical and systematic
limitations of the training data through a neural network. In addition
to the usual output the Bayesian network provides event-by-event error
bars.  A limitation we encounter in Ref.~\cite{ours} is that forcing
the network output onto a closed interval, like a probability $p \in
[0,1]$, strongly correlates the the central value and the error bars
in the network output. This makes it difficult to track systematic
uncertainties.

We circumvent this problem by extracting the transverse momentum,
which does not live on a closed interval. In the previous section this
allowed us to decompose $\sigma_\text{tot}$ into a statistical
component, $\sigma_\text{pred}$, and a systematic component,
$\sigma_\text{stoch}$.  What we still need to study is the actual
output distribution of the Bayesian network, $p(p_T| M)$, and how it
compared to the truth information from the test data.

\subsubsection*{Variance of training data and $\sigma_\text{stoch}$}

\begin{figure}[t]
\centering
\includegraphics[width=0.325\textwidth]{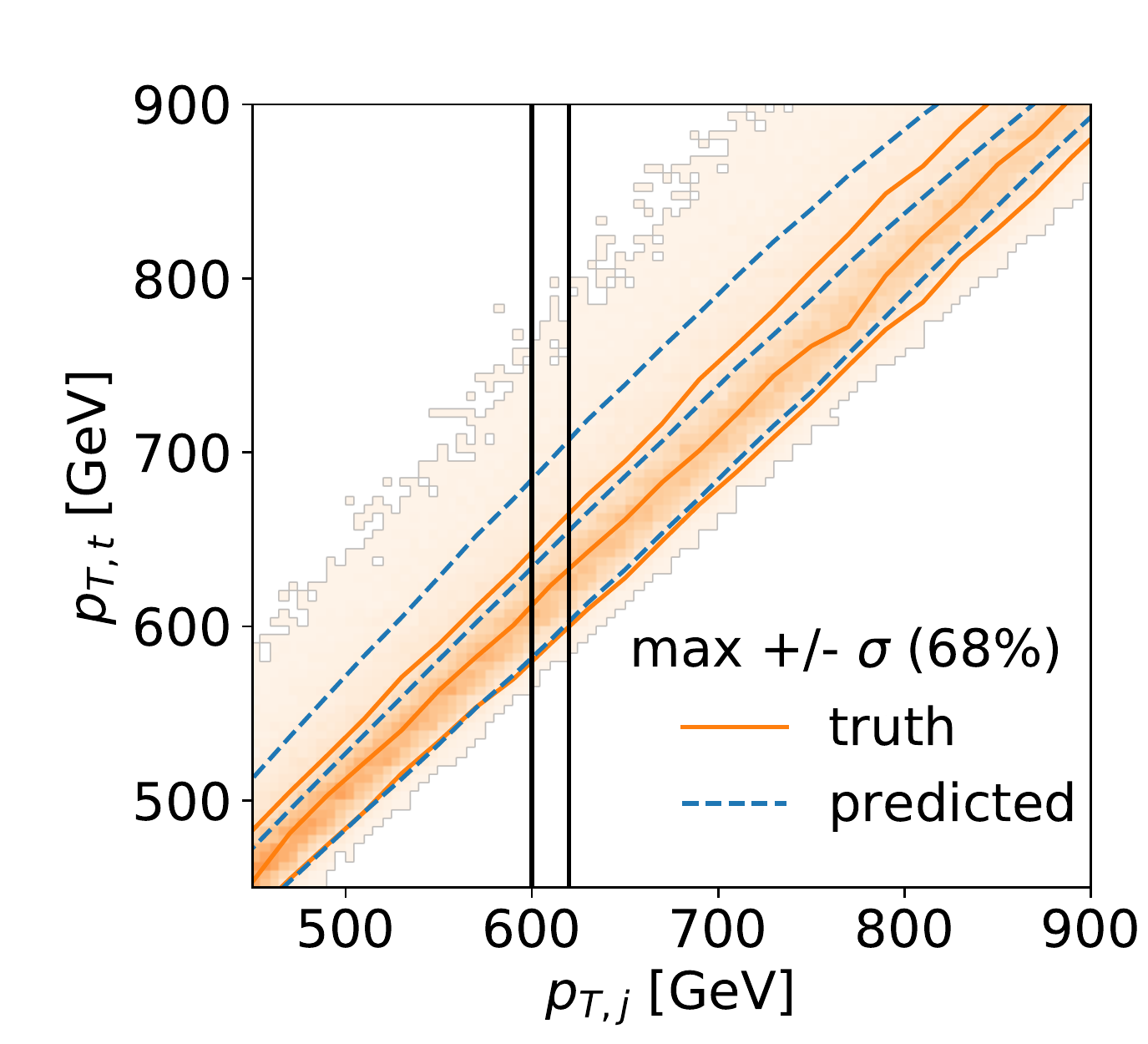}
\includegraphics[width=0.325\textwidth]{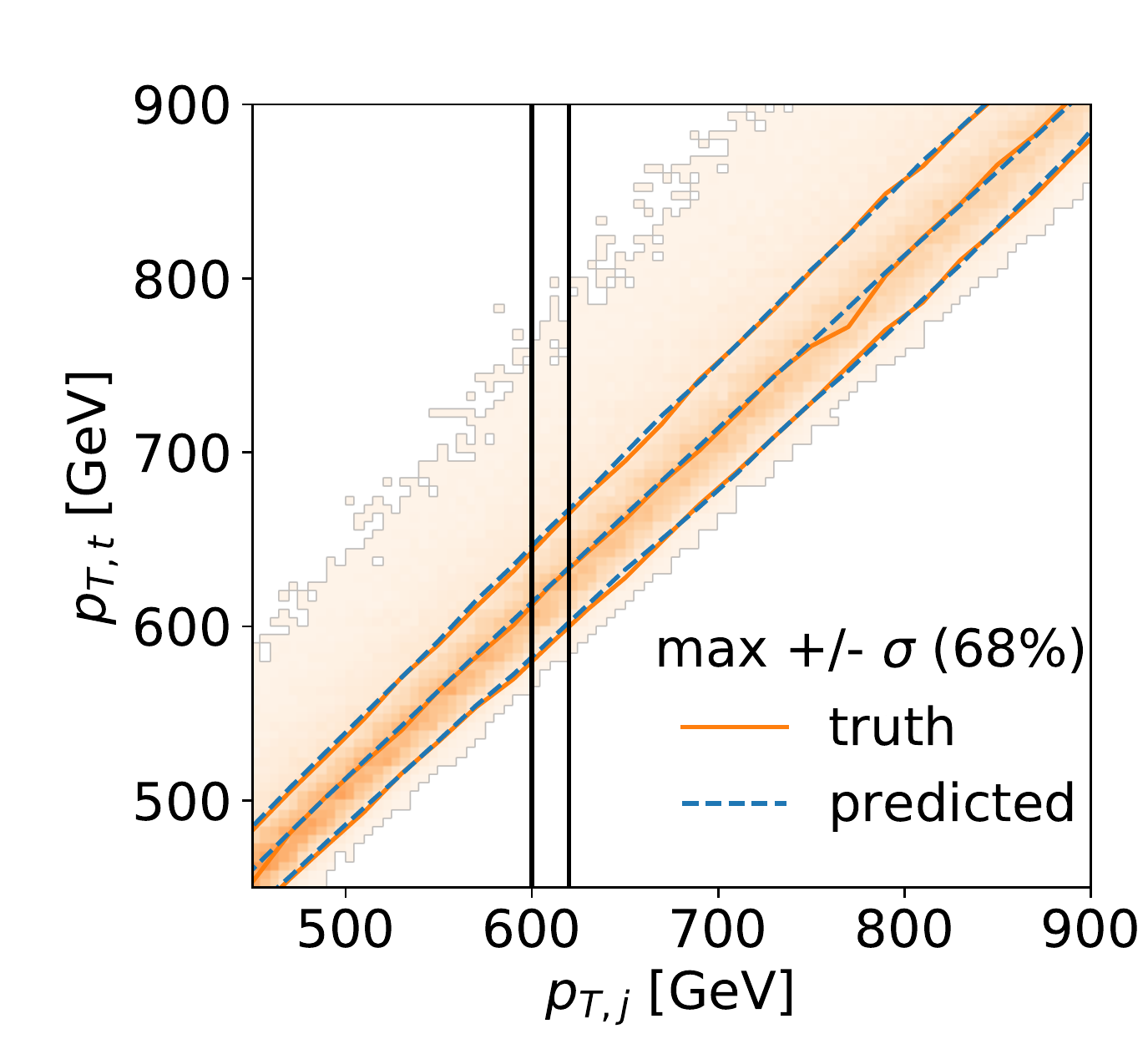}
\includegraphics[width=0.325\textwidth]{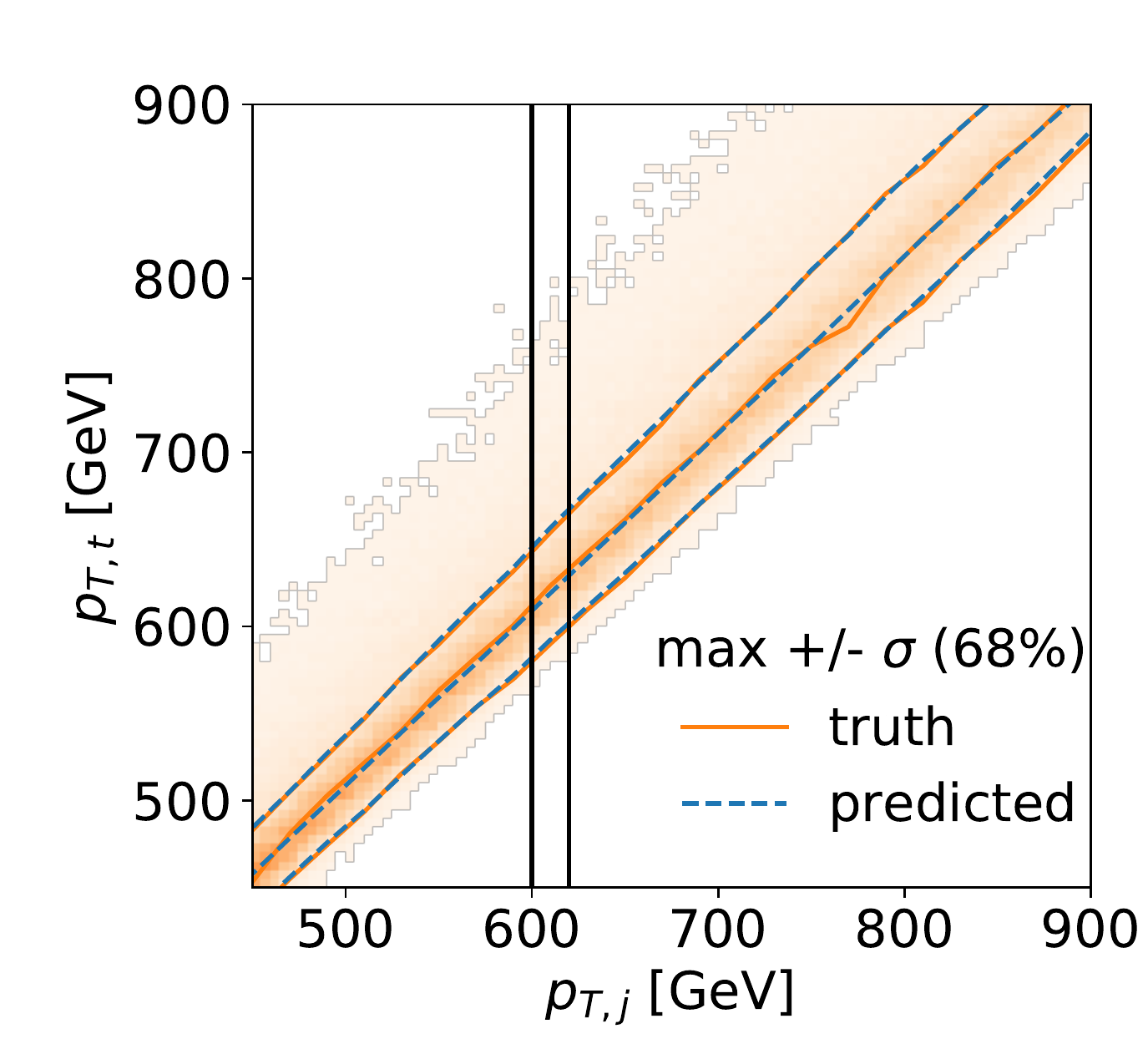}
\includegraphics[width=0.325\textwidth]{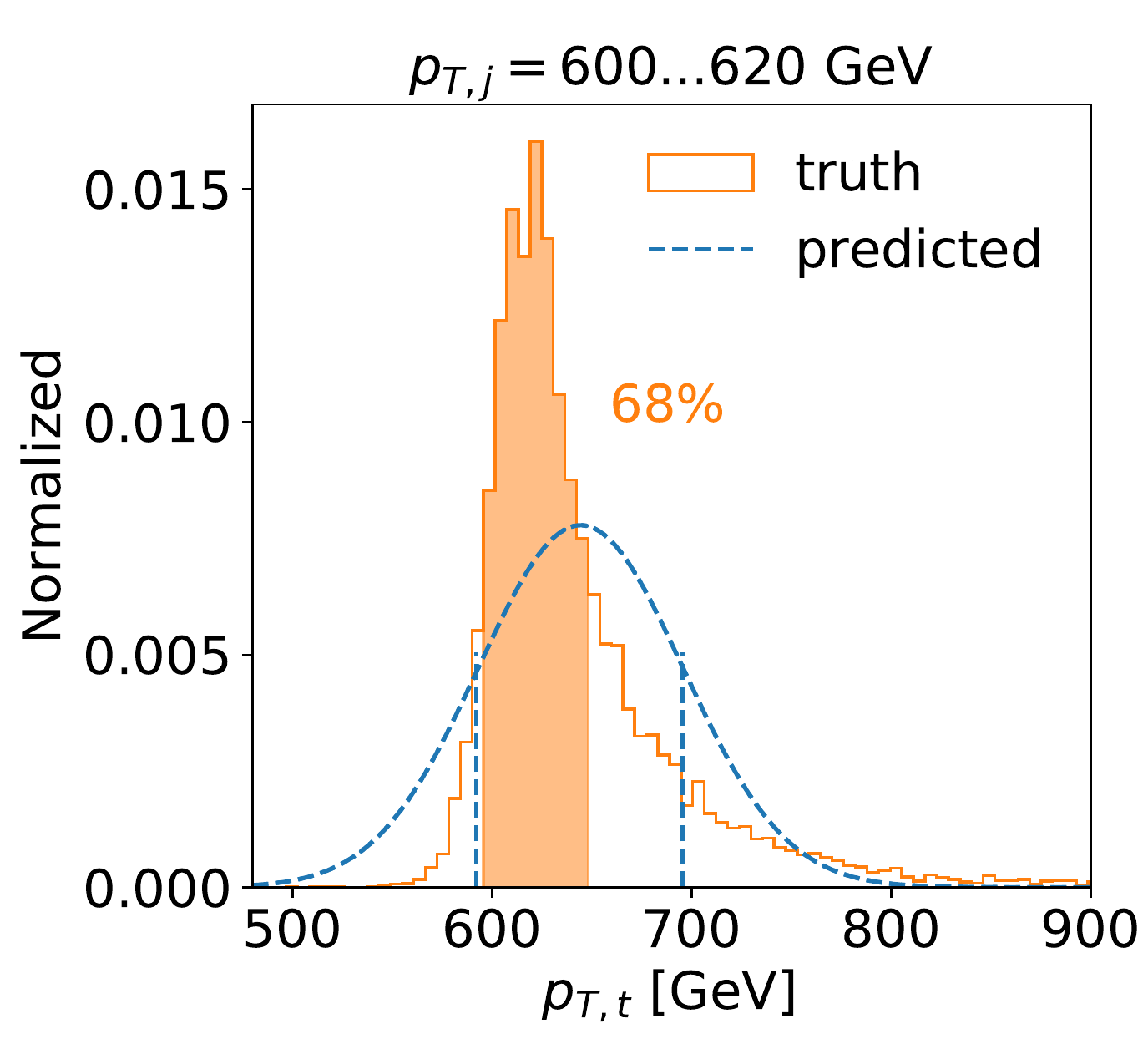}
\includegraphics[width=0.325\textwidth]{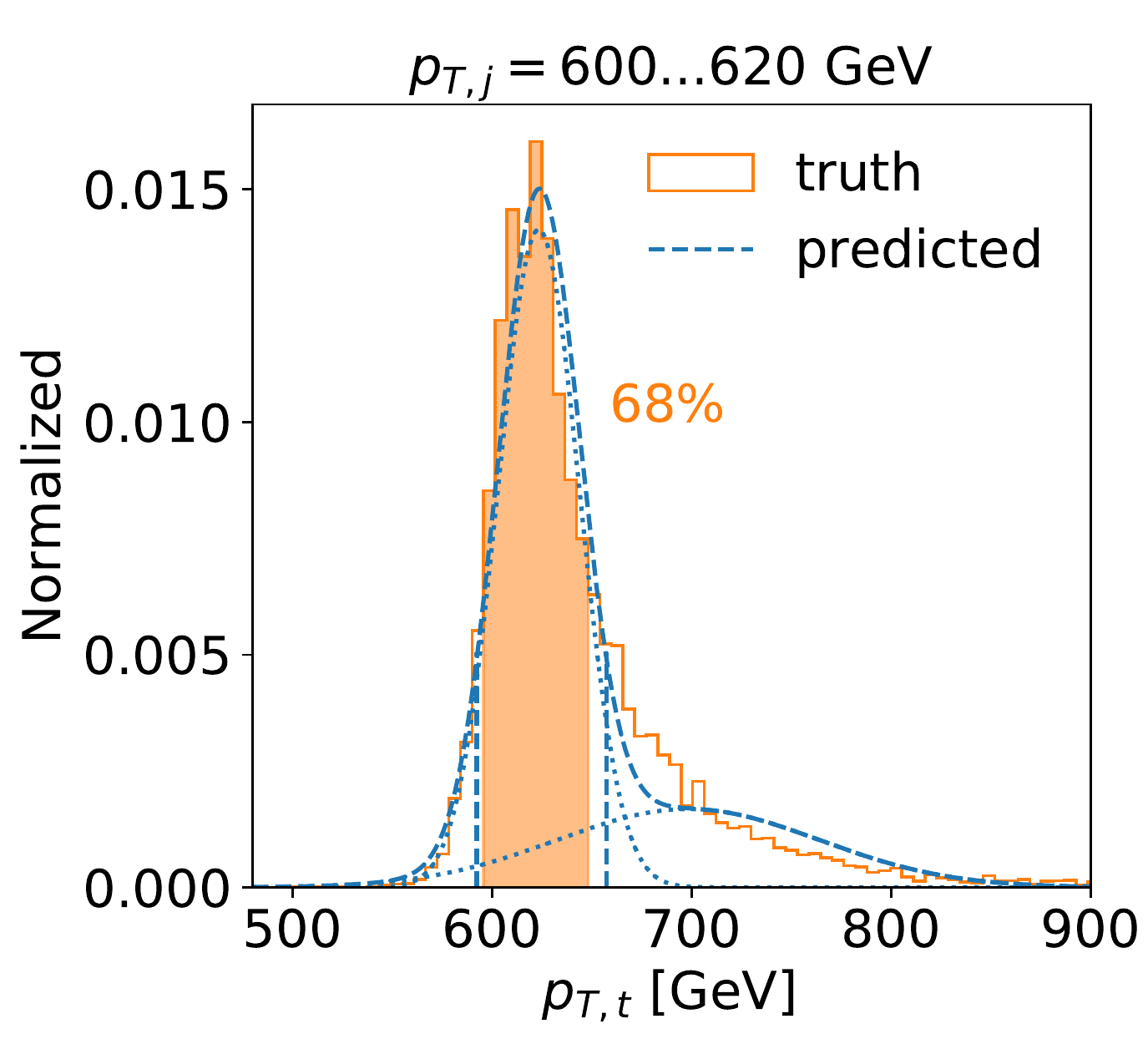}
\includegraphics[width=0.325\textwidth]{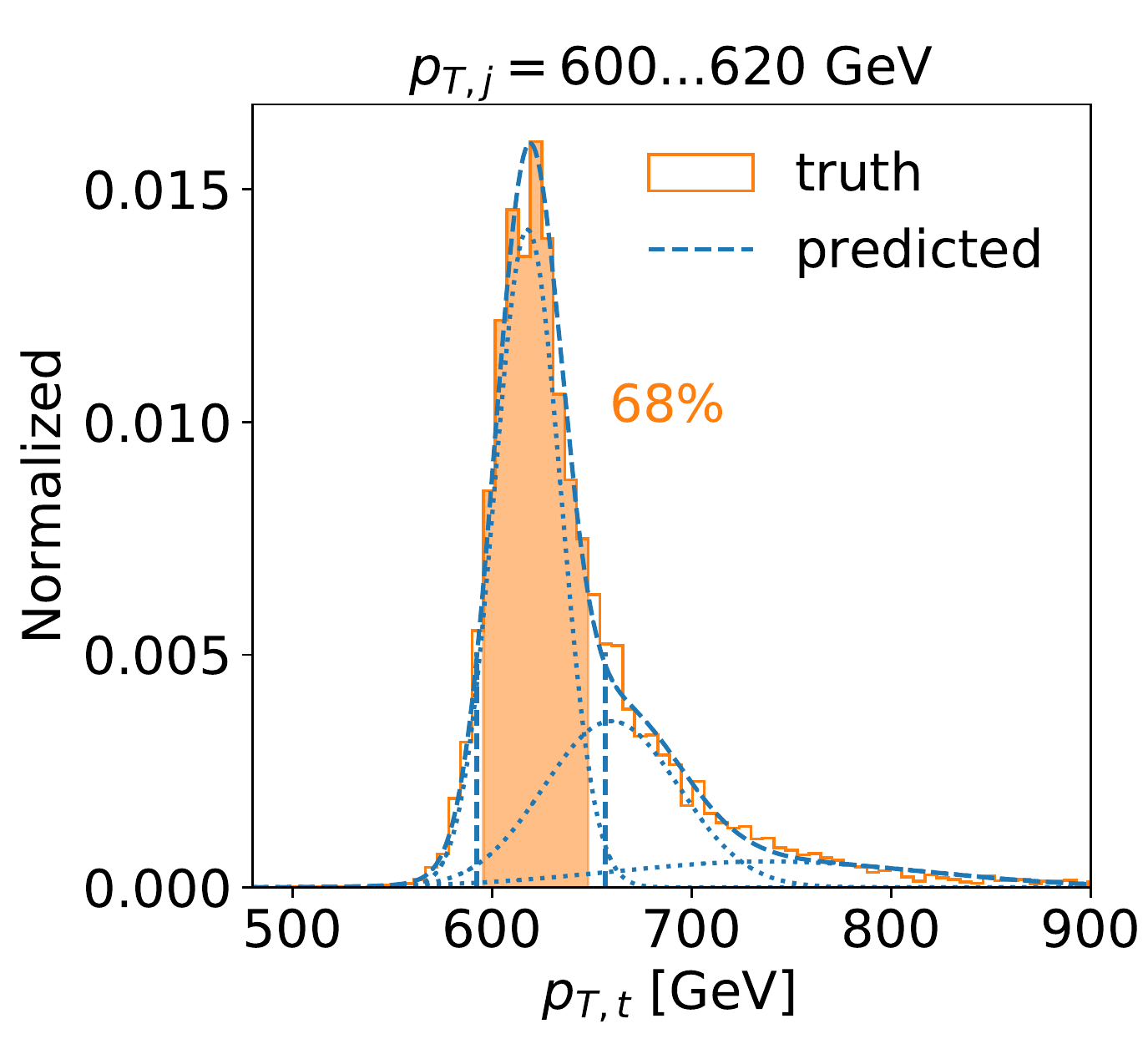}
\caption{Upper: 2-dimensional distribution of $\ptt$ vs $p_{T, j}$
  including its 68\%~CL around the maximum. In blue we show the BNN
  results. Lower: $\ptt$-distribution for a narrow slice in $p_{T,
    j}$. From left to right we approximate $\ptt$ with one, two, and
  three Gaussians.}
\label{fig:intrinsic_unc}
\end{figure}

In the upper left panel of Fig.~\ref{fig:intrinsic_unc} we show the
correlation of $\ptt$ and $p_{T, j}$.  The orange curves represent the
maximum and the $68\%$~CL interval in 20~GeV bin.  The corresponding
maximum and $68\%$~CL interval of the BNN output are illustrated in
blue. Both confidence intervals are constructed by requiring equal
functional values at both ends.  In the lower left panel we see why
the two sets of curves agree very poorly: for the narrow $p_{T,j}$
slide the $\ptt$ distribution is all but Gaussian, while the
Bayesian output in our naive approach is forced to be Gaussian, as
seen in Eq.\eqref{eq:output}.

From Sec.~\ref{sec:bayes} we know that it is not necessary to assume
that the Bayesian network output is Gaussian. As a simple
generalization we can replace the two-parameter Gaussian form of
$p(M|\omega)$ in Eq.\eqref{eq:kl_loss1} with a mixture of Gaussians,
\begin{align}
p(M| \omega) = \sum_i \alpha_{i, \omega} \; G (\langle p_T \rangle^{(i)}_\omega, \sigma^{(i)}_{\text{stoch}, \omega}) \, ,
\end{align}
with $\sum_i \alpha_{i, \omega}=1$. The network output from Eq.\eqref{eq:output}
then becomes
\begin{align}
\text{NN} (\omega) = 
\begin{pmatrix}
\alpha_{1, \omega} & \alpha_{2, \omega} & \cdots \\
\langle p_T \rangle_{\omega}^{(1)} & 
\langle p_T \rangle_{\omega}^{(1)} & \cdots \\
\sigma_{\text{stoch}, \omega}^{(1)} &
\sigma_{\text{stoch}, \omega}^{(2)} & \cdots \\
\end{pmatrix} 
\label{eq:output_multi}
\end{align}
To guarantee $\sum_i \alpha_{i, \omega}=1$ we use SoftMax as an
activation function for $\alpha_{i, \omega}$ and the SoftPlus function
for $\sigma_{\text{stoch}, \omega}^{(i)}$ to ensure positive values.
In the center and right sets of panels in Fig.~\ref{fig:intrinsic_unc}
we see what happens if we use two or three Gaussians, specifically
with the parameters averaged over weights and jets in a bin. For three
Gaussians the BNN output and the $\ptt$ distribution agree
perfectly. The corresponding parameters are shown in
Tab.~\ref{tab:gm_mixture}.

\begin{table}
\centering
\begin{small}
\setlength{\tabcolsep}{4.2pt} 
\begin{tabular}{c|ccc|cc|cc|ccc}
\toprule
& $\alpha^{(i)}$ & $\langle p_T \rangle^{(i)}$ & $\sigma_\text{stoch}^{(i)}$ 
& $\sigma_\text{stoch}$ & $\mse$ 
& $\langle p_{T, t} \rangle$ & $\langle \ptt \rangle$ 
& Max & 68\%CL & 68\%CL (truth) \\
\midrule
1 & 1 & 644.4 & 51.43 
  &  51.4 &  
  & 644.4 &  
  & 644.4 & 593.0...695.9 \\
\midrule
2 & $\begin{matrix} 0.72 \\ 0.28 \end{matrix} $
  & $\begin{matrix} 623.4 \\ 698.3 \end{matrix} $
  & $\begin{matrix} 20.4 \\ 65.6 \end{matrix} $
  & 51.1 &  
  & 644.1 &  
  & 623.4 & 592.4...657.3 \\
\midrule
3 & $\begin{matrix} 0.59 \\ 0.30 \\ 0.11 \end{matrix} $
  & $\begin{matrix} 617.8 \\ 659.8 \\ 738.6 \end{matrix} $
  & $\begin{matrix} 16.6 \\ 33.7 \\ 78.6 \end{matrix} $
  & 51.5 & 52.2
  & 643.8 & 643.8 
  & 619.1 & 592.4...656.8 
  & 590.0...654.0 \\
\bottomrule
\end{tabular} 
\end{small}
\caption{Parameters used in Fig.~\ref{fig:intrinsic_unc}, specifically
  $p_{T, j} = 600...620$~GeV.}
\label{tab:gm_mixture}
\end{table}

Technically, we follow Sec.~\ref{sec:bayes} in extracting
$\sigma_\text{stoch}$ and $\sigma_\text{pred}$ independently of the
form of the underlying assumption.  Two aspects render this
computation slightly expensive: the integration over all weights and,
if required, the combination of different predictions in one $p_{T,j}$
bin. On the other hand we know that $\sigma_\text{pred} \ll
\sigma_\text{stoch}$ and we can always use narrow bin sizes. This
means that in both cases we can replace the integrals by simply
averaging over the parameters of the Gaussian mixture model.  This
implementation is computationally less expensive and gives us simple
analytic expressions from which we extract the maximum and $68\%$~CL
interval.

\subsubsection*{Noisy labels}

A crucial question in experimental physics is how we include a
systematic uncertainty for instance on the jet energy scale in the
training procedure. We can understand such an energy calibration when
we remind ourselves that the jets in the calibration sample come with
a measured reference value for their energies and the corresponding
error bar; and that the calibration sample in our case is the training
sample. There are two ways we can include the error on the calibration
measurements in our analysis:
\begin{enumerate}
\setlength\itemsep{-0.3em}
\item[1A.] fix the label or `true energy' and smear the jets in the training sample;
\item[1B.] fix the jets and smear the continuous label in the training sample;
\item[2.] train the Bayesian network on the smeared label-jet combination;
\item[3.] extract a systematics error bar for each jet in the test sample.
\end{enumerate}
In Ref.~\cite{ours} we have followed the option 1A and encountered
some practical/numerical problems when tracing the corresponding
systematics to the network output. In this study we shift to the less
standard and yet straightforward option 1B. We assume that jet
calibration incorporates external information on the training sample,
be it another measurement or a theory requirement (one-shell
$Z$-decays) or a MC prediction. This information defines a label
together with a corresponding error bar. This means we train our
network on a fixed sample of jets with a smeared label representing
the full reference measurement. In this approach we can trivially
include additional uncertainties from pre-processing the training
data, like running a jet algorithm of the $Z$-sample, removing
underlying event and pile-up, etc.  As a side effect our setup also
allows us to capture possible transfer uncertainties, whenever our
test sample cannot easily be linked to the training sample. In the ML
literature such uncertainties are referred to as out-of-sample error.

\begin{figure}[t]
\centering
\includegraphics[width=0.48\textwidth]{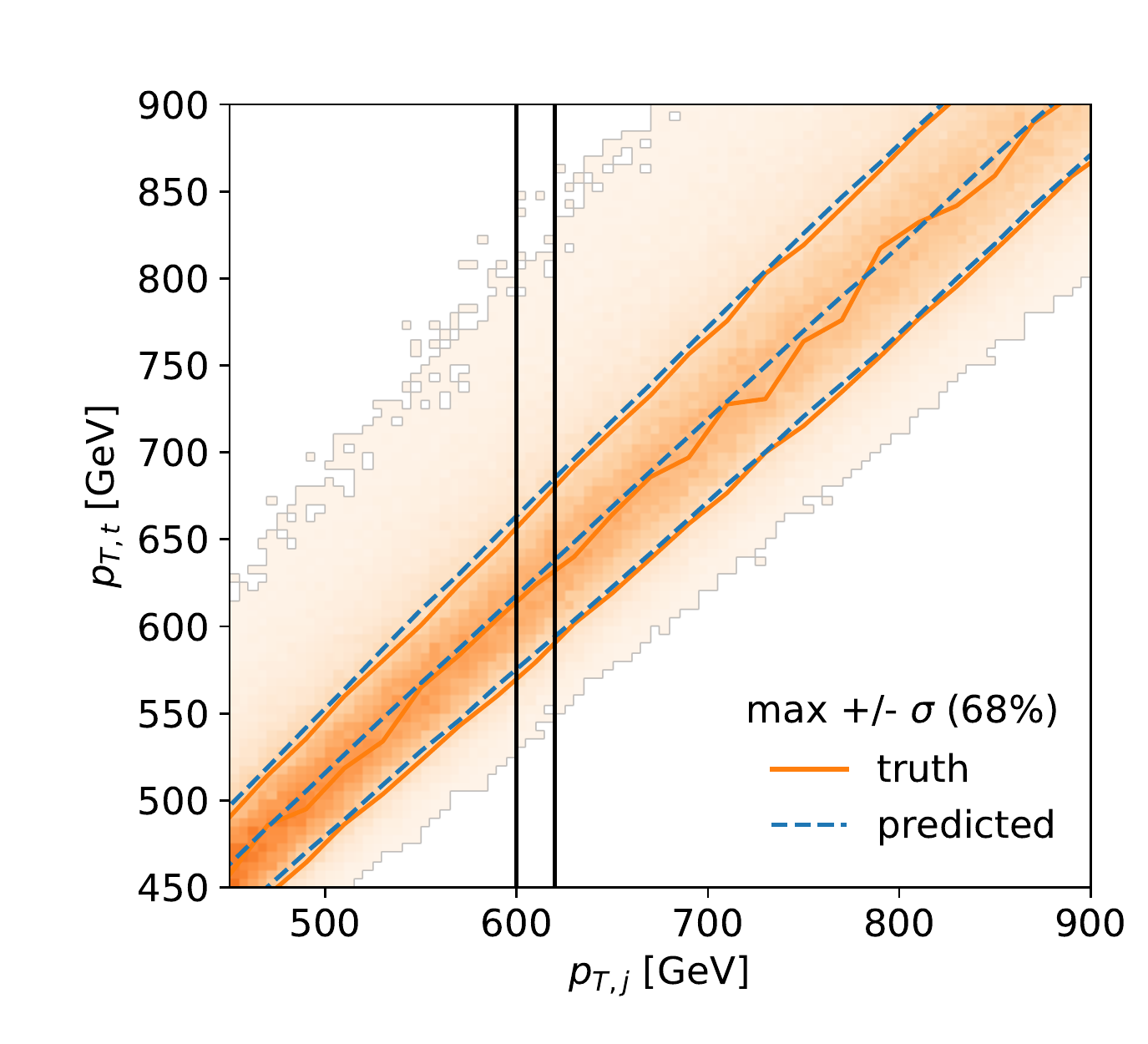}
\hspace*{0.01\textwidth}
\includegraphics[width=0.48\textwidth]{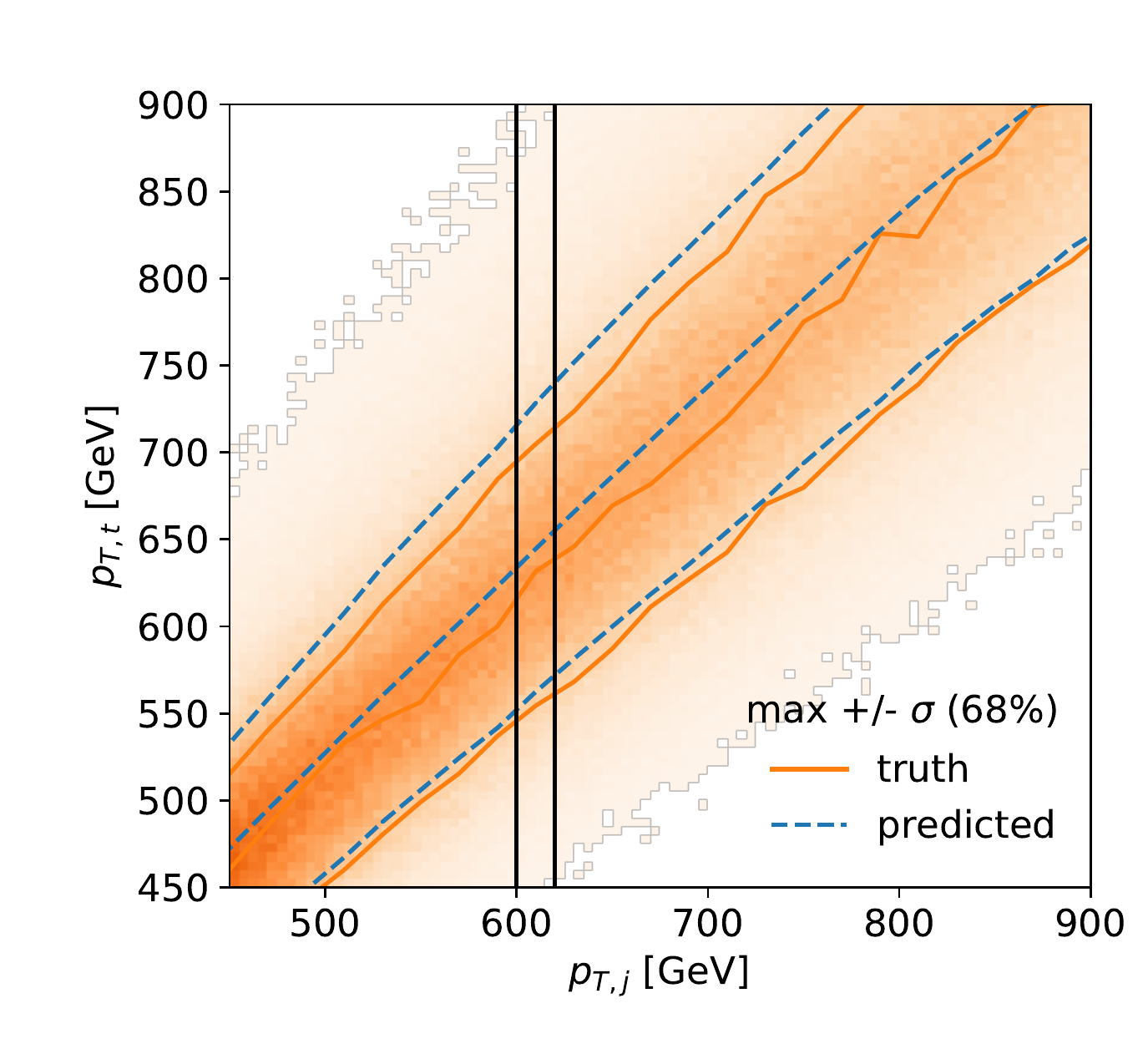} \\
\includegraphics[width=0.48\textwidth]{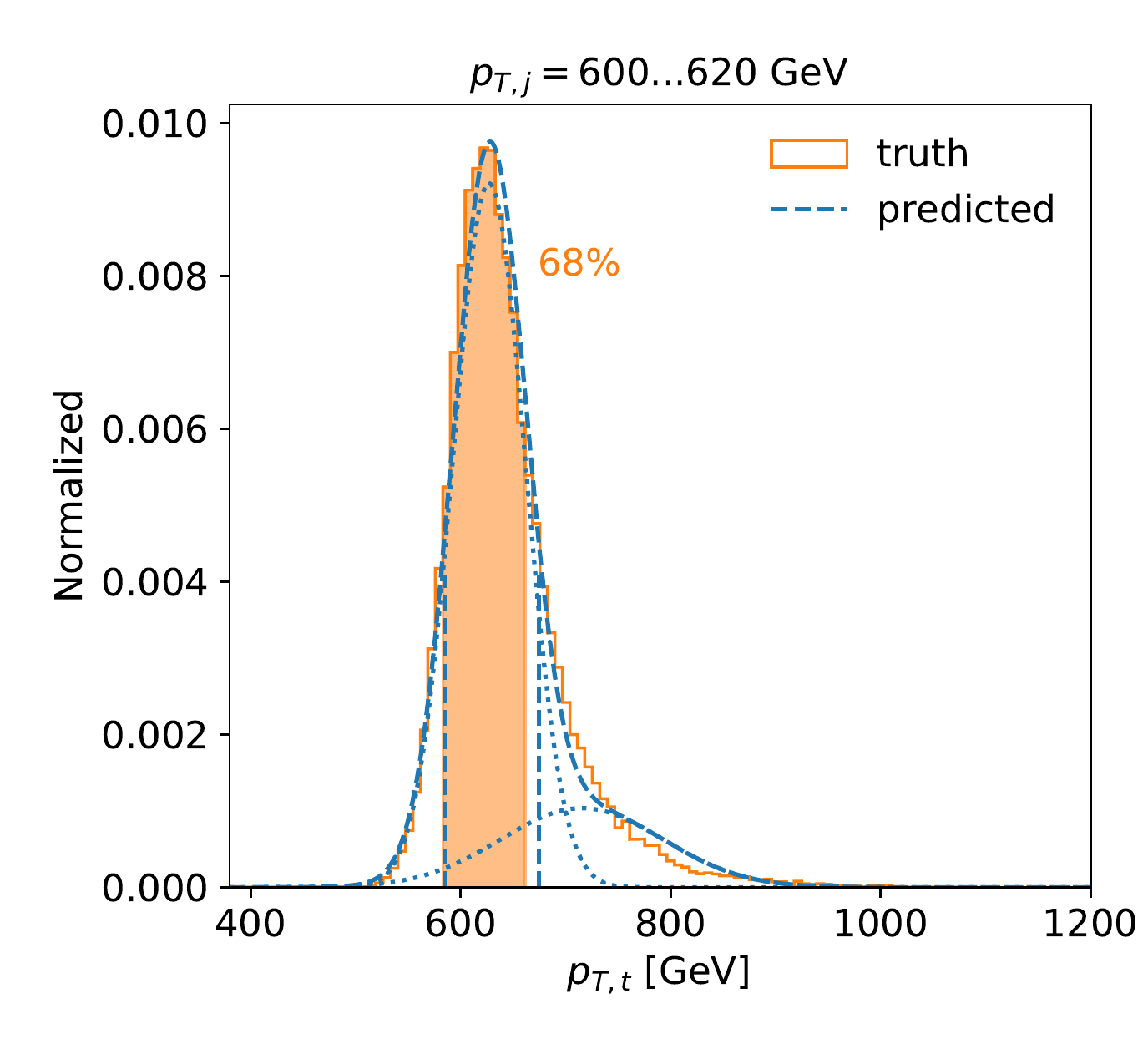}
\hspace*{0.01\textwidth}
\includegraphics[width=0.48\textwidth]{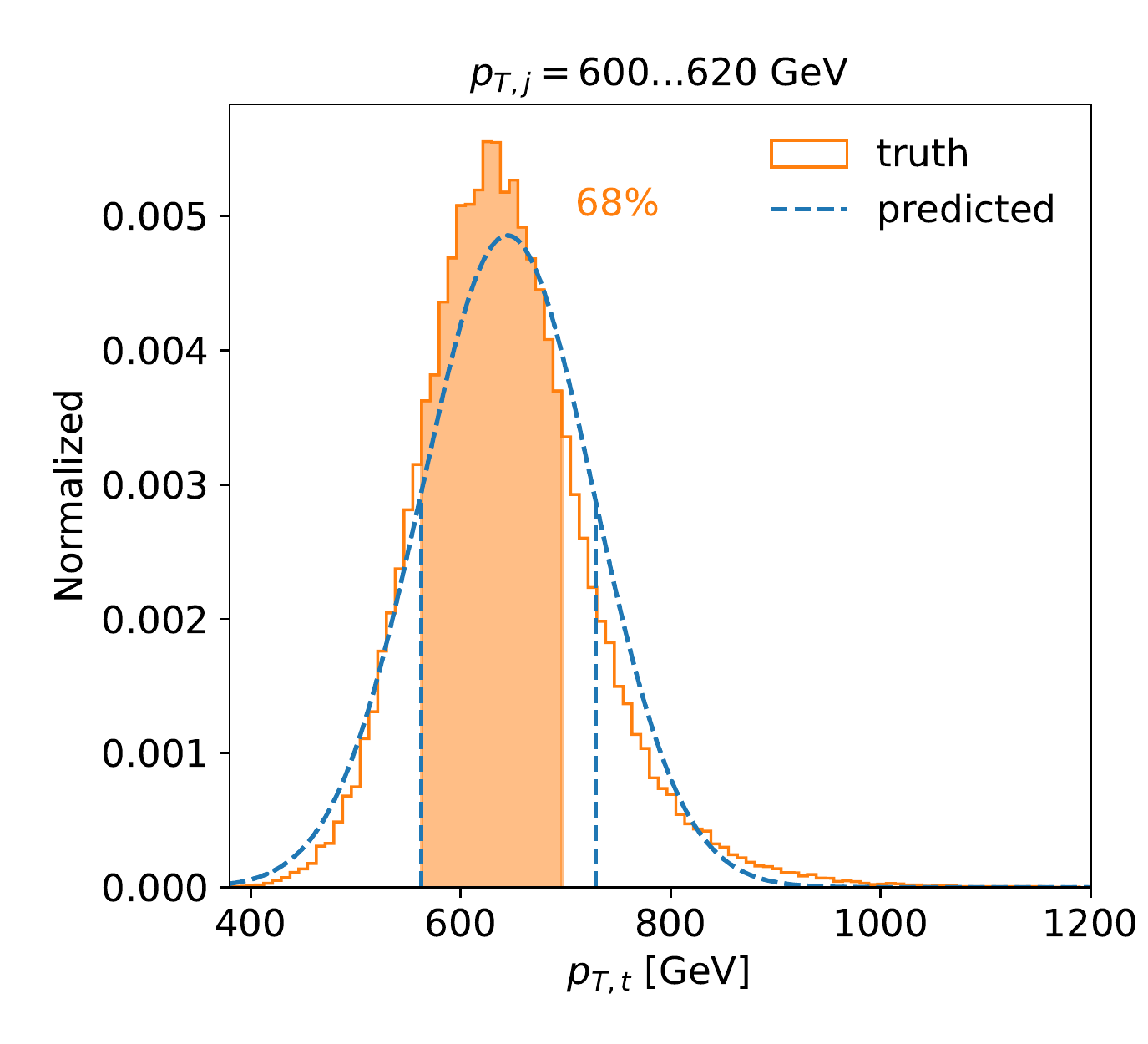}
\caption{Upper: 2-dimensional distribution of $\ptt$ vs $p_{T,j}$
  including its 68\%~CL around the maximum, after adding 4\% (left)
  and 10\% (right) Gaussian noise on the top momentum label.  In blue
  we also show the BNN error estimate. Lower: corresponding
  $\ptt$-distribution for a narrow slice in $p_{T,j}$.}
\label{fig:smearing_labels_1}
\end{figure}

To illustrate and test our setup we smear $\ptt$, the label in the
training data, according to Gaussians with widths of 
\begin{align}
\sigma_\text{smear} = (4~...~10) \% \times \ptt \; .
\end{align}
In Fig.~\ref{fig:smearing_labels_1}.  we see that for a small amount of
smearing the non-Gaussian shape of Fig.~\ref{fig:intrinsic_unc}
remains, so we use two Gaussians in the BNN. For sizeable Gaussian
smearing we see that the resulting distributions all assume a Gaussian
shape and we can stick to the single-Gauss standard BNN. In both cases
the distribution of the BNN output and the (smeared) label $\ptt$
agree almost perfectly.

\begin{figure}[t]
\centering
\includegraphics[width=0.48\textwidth]{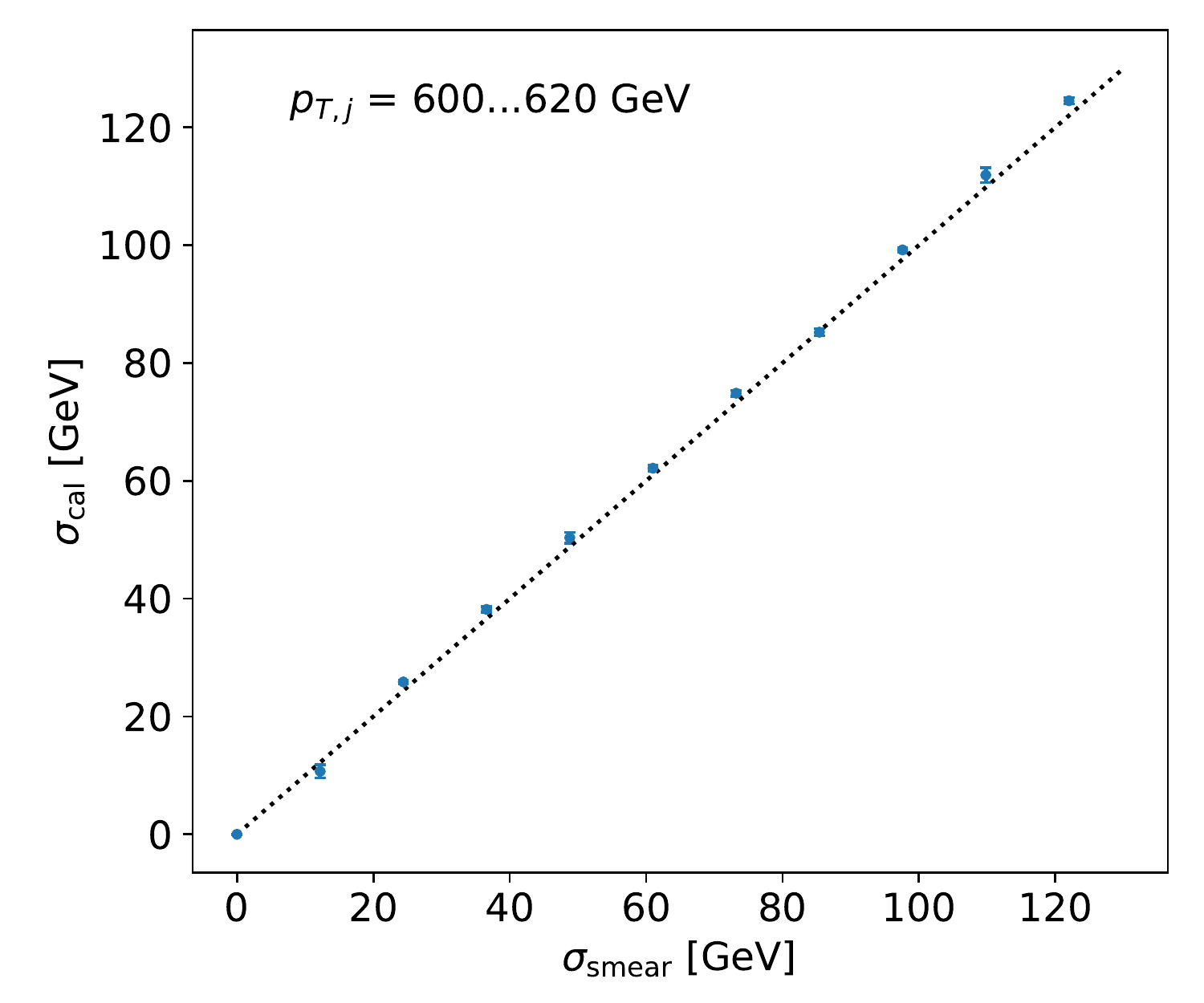}
\caption{Correlation between $\sigma_\text{stoch}$, as given by the
  Bayesian network and the smearing $\sigma_\text{smear}$ applied to
  the label in the training data. The baseline
  $\sigma_{\text{stoch},0}$ is defined as $\sigma_\text{stoch}$ in the
  limit of no smearing. The error bars indicate the standard deviation
  from five independent trainings.}
\label{fig:smearing_labels_2}
\end{figure}

From the previous sections we know that the reported uncertainty by
the BNN includes a statistical uncertainty vanishing with an
increasing amount of training data and a systematic uncertainty
representing the stochastic nature of the training data. When we
introduce another uncertainty induced by smeared labels we expand
Eq.\eqref{eq:def_sigmas} to
\begin{align}
\sigma_\text{tot}^2
&= \sigma_\text{stoch}^2
+ \sigma_\text{pred}^2 \notag \\
&= \sigma_{\text{stoch},0}^2
+ \sigma_\text{cal}^2  
+ \sigma_\text{pred}^2 
\quad \Leftrightarrow \quad 
\sigma_\text{cal}^2 
= \sigma_\text{stoch}^2
- \sigma_{\text{stoch},0}^2 \; ,
\end{align}
added in quadrature because of the central limit theorem. The baseline
value $\sigma_{\text{stoch},0}$ is defined as $\sigma_\text{stoch}$ in
the limit of no smearing. In Fig.~\ref{fig:smearing_labels_2} we show
how $\sigma_\text{cal}$ correlates with the input
$\sigma_\text{smear}$ over a wide range of scale uncertainties.  As
usually, the error bar represents the standard deviation from five
independent trainings. This correlation shows that our network picks
up the systematic uncertainties from smeared training labels
perfectly. We note that, as before, this analysis does not require a
Gaussian shape of the network output.


\section{Outlook}

We have shown that Bayesian networks keep track of statistical and
systematic uncertainties in the training data and translate them into
a jet-by-jet error budget for instance in a momentum
measurement. Outside particle physics it is not unusual to treat
uncertainties as a smearing of labels, whereas in particle physics we
usually model them by smearing the input data. We show that smearing
labels is a natural, feasible, and self-consistent strategy in
combination with deep learning. An advantage of this approach is that
the treatment of uncertainties is moved from the evaluation time to
the training time and so-trained networks accurately report
predictions of the central value as well as systematic
uncertainties. 

We have shown that the corresponding Bayesian networks allow us to
cleanly separate statistical and systematic uncertainties. In
addition, the smeared labels are ideally suited to translate
uncertainties from reference or calibration data to the network
output.

Technically, we have modified the Bayesian network approach of
Ref.~\cite{ours} to include non-Gaussian behavior. This step is
crucial for modeling systematic uncertainties in general.

We emphasize that before this approach can be generally adapted,
open questions such as multiple correlated uncertainties and the
translation between input-uncertainties and label-uncertainties need
to be answered. However, our first results show great promise for
smeared labels describing uncertainties in particle physics
applications of deep learning.\bigskip

\begin{center} \textbf{Acknowledgments} \end{center}

We would like to thank Ben Nachman for many very useful discussions
and Manuel Hau{\ss}mann for getting us into Bayesian neural networks.
ML is funded through the Graduiertenkolleg \textsl{Particle Physics
  Beyond the Standard Model} (GRK 1940). The authors acknowledge
support by the state of Baden-W\"urttemberg through bwHPC and the
German Research Foundation (DFG) through grant no INST 39/963-1
FUGG(bwForCluster NEMO).  GK is acknowledges support by the Deutsche
Forschungsgemeinschaft under Germany's Excellence Strategy - EXC 2121
\textsl{Quantum Universe} - 390833306.

\appendix
\section{Comparison to smeared data}

\begin{figure}[h!]
\centering
\includegraphics[width=0.48\textwidth]{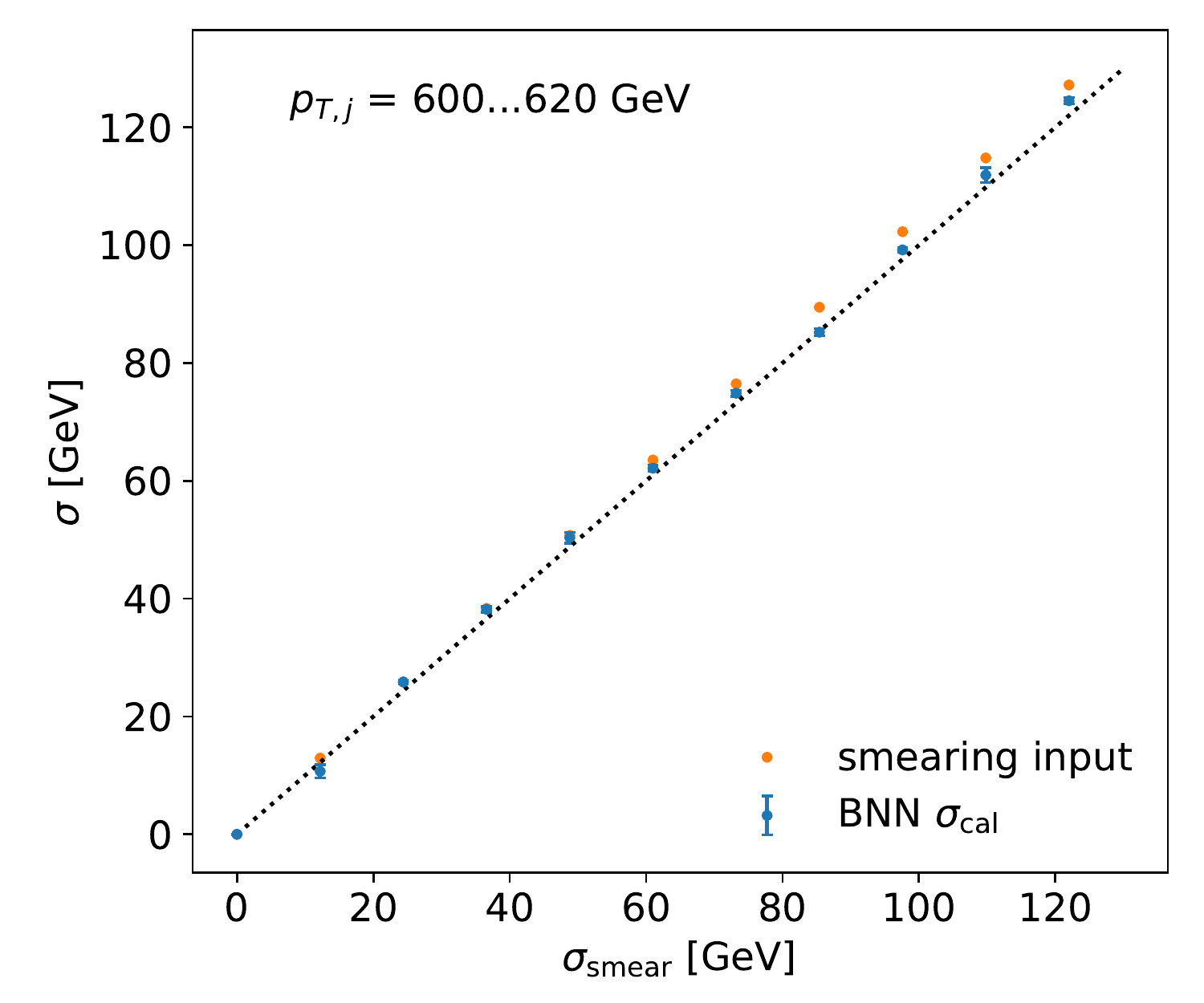}
\caption{Comparison of the Bayesian approach, taken from
  Fig.~\ref{fig:smearing_labels_2} (blue), and smearing of input data
  (orange). When smearing the input data, we train a Bayesian network
  on nominal events and test it on inputs modified up and down by
  $\sigma_\text{smear}$.}
\label{fig:smearing_data}
\end{figure}

To further validate the proposed approach,
Fig.~\ref{fig:smearing_data} compares the performance of the BNN
approach with a more traditional smearing of the input objects. For
smearing the objects we use a Bayesian neural network trained on data
without smearing and evaluate this network on a test dataset with
modified inputs.  Each jet in the test sample is smeared once up and
once down, then the difference of the two network outputs is evaluated
and divided by two. We then show the average in the given $p_{T,
  j}$-range.  The BNN prediction is in good agreement with modified
inputs, giving additional confidence in uncertainty predicted by the
Bayesian network.

\bibliography{refs}

\end{document}